\documentclass[
prd,
superscriptaddress,
10 pt,
preprintnumbers,
notitlepage,
secnumarabic,
nobibnotes,
nofootinbib,
showpacs
]{revtex4-1}

\usepackage[T1]{fontenc}
\usepackage{bm}
\usepackage[pdftex]{color,graphicx}
\usepackage{amssymb}
\usepackage{mathtools}
\usepackage{tabularx}
\usepackage{dsfont}
\usepackage[artemisia]{textgreek}
\usepackage{hyperref}
\usepackage{xcolor}
\usepackage{enumerate}
\usepackage{hyperref}

\usepackage{perpage}
\MakePerPage{footnote}

\DeclareMathAlphabet\mathbfcal{OMS}{cmsy}{b}{n}
\newcommand{\kbar}{\mathchar'26\mkern-9mu k}
\renewcommand{\i}{\mathrm{i}}

\begin{document}
\title{Critical Insight into the Cosmological Sector of Loop Quantum Gravity}

\author{Jakub Bilski}
\email{bilski@zjut.edu.cn}
\affiliation{Institute for Theoretical Physics and Cosmology, Zhejiang University of Technology, 310023 Hangzhou, China}

\author{Antonino Marcian\`o}
\email{marciano@fudan.edu.cn}
\affiliation{Center for Field Theory and Particle Physics \& Department of Physics, Fudan University, 200433 Shanghai, China}


\begin{abstract}
\noindent
This article sheds new light on the problem of cosmological reduction in Loop Quantum Gravity. We critically analyze Quantum Reduced Loop Gravity --- an attempt to extract the cosmological sector of the full theory. We reconsider the reduction procedure applied to the states of the kinematical Hilbert space, developing a comparative analysis with previous efforts in the literature. We show that the constraints of the model were formerly instantiated in an inconsistent fashion, leading to an overconstrained dynamics and an ill-defined Hilbert space. We then scrutinize alternative implementations of symmetry-reduction. While remaining unaffected by the shortcomings encountered in Quantum Reduced Loop Gravity, these latter procedures bridge the gap between the full theory and former endeavors in Loop Quantum Cosmology.

\end{abstract}
\maketitle

	
\section{Introduction}\label{I}
\noindent 
Despite over the last three decades optimistic acclamation was often bestowed upon Loop Quantum Gravity (LQG) \cite{Thiemann:2007zz}, no clear resolution of the quantization of the Hamiltonian constraint problem was sorted out. We therefore still ignore the physical Hilbert space of the theory, and consequently its ground state. The very same structure of the vacuum, which unfortunately is still very far from being tackled within the theory, is suggested to be non-trivial by considerations based on non-abelian gauge theories\footnote{The well known non-trivial structure of the vacuum was emphasized for gauge theories, providing a possible framework to account for quantum theories of gravity, by inspections of their instantonic solutions, as argued in Ref.~\cite{Addazi:2017xur}.}. Already two decades ago, the overall situation of incompleteness eventually urged to grasp more insights by a development of a parallel simpler theory, Loop Quantum Cosmology (LQC) \cite{Ashtekar:2003hd,Bojowald:2008zzb,Ashtekar:2011ni}, established to deal with the quantization of the symmetry-reduced phase space of the full theory, LQG. However, the link between LQG and LQC is neither obvious nor obviously able to shed light on the quantization issues, as the longstanding lack of solutions to the former, despite the development of LQC, proves. Other approaches then naturally followed, often motivated by the purpose of linking LQC to the full theory of LQG, since quantization and symmetry-reduction need not, \textit{a priori}, to commute. Several possibilities were investigated within the literature \cite{Engle:2007zz,Brunnemann:2007du,Engle:2013qq,Bodendorfer:2015hwl}, trying to unravel how the quantum configuration spaces of LQC can be embedded into the full theory. Lights on the use of spinfoam techniques was sought in \cite{Rovelli:2009tp}, while coherent state techniques were proposed within the Group Field Theory approach in \cite{Gielen:2013naa,Oriti:2016qtz}.

Quantum reduced loop gravity (QRLG) is chronologically one of the latest attempts, developed in \cite{Alesci:2013xd, Alesci:2013xya, Alesci:2014uha,Alesci:2017kzc} --- for a review see \cite{Alesci:2016gub}. It relies on imposing weak gauge-fixing conditions to the states of the kinematical Hilbert space of the full theory, LQG. This peculiarity was argued to allow recovering the cosmological sector directly from LQG. Classically, the gravitational systems considered are those ones described by dreibein fields gauge-fixed to a diagonal form. The gauge-fixing conditions are then applied weakly on the kinematical Hilbert space of the full theory. As a result, Bianchi I models were thought to be successfully recovered within the framework --- see \cite{Ashtekar:2009vc} for a description of the Bianchi I extension to LQC. Furthermore, there are studies seeking to show that within the semiclassical limit, QRLG reproduces the effective Hamiltonian of LQC \cite{Alesci:2014uha,Alesci:2014rra}, in the $\mu_0$ regularization scheme. It has been also claimed that the effective improved dynamics proposed in \cite{Ashtekar:2006wn}, can be inferred in this framework by averaging over the ensemble of the classically equivalent states \cite{Alesci:2016rmn}.

For the aforementioned reasons, QRLG was conjectured to provide a novel derivation of earlier results of LQC, including the realization of the singularity-resolution scenario. But disregarding the emanation of LQG, the full theory will remain unsolved until the Hamiltonian constraint problem will be solved, and matter fields will be taken into account. Within the framework of QRLG, since this introduces a graph structure underlying the description of the continuous universe at the classical level, and since the origin of the discretization must be recovered at the quantum level, the quantization of the matter fields shall be achieved via the same tools of LQG \cite{Thiemann:1997rq,Thiemann:1997rt}. This was providing encouragement that QRLG might have offered a framework to test the implications of the loop quantization of matter fields, as first suggested by the analysis of a scalar matter field in \cite{Bilski:2015dra}, and then by the implementation of gauge vector fields in \cite{Bilski:2016pib,Bilski:2017dgi} and related applications.

QRLG initially appeared as a promising way to reconcile a cosmological toy model, LQC, while a full theory remained under construction. But this did not come without its flaws. Actually, as we will question throughout the paper, the shortcomings of the model are enough to state its internal inconsistency. Specifically, we will argue about the fate of the theory, re-examining the reduction procedure applied to the states of the kinematical Hilbert space of LQG, and developing a comparative analysis with previous attempts formulated in the literature of QRLG, while seeking to unravel the cosmological sector of the full theory. We show that constraints are inconsistently implemented within this framework, leading to an overconstrained dynamics.

We will elaborate on the methods to attain a proper definition of quantum states of the kinematical Hilbert space, spelling arguments hinging toward an alternative implementation of the symmetry-reduction procedure. While remaining unaffected by the flaws of QRLG, this new procedure appears to be the first --- up to our knowledge --- consistent example of a bridge between the full theory and the other attempts belonging to the LQC scenario. Following the standard procedures of the reduced phase space quantization, we then demonstrate how the canonical variables of LQG are simplified exactly to the form of the anisotropic extension of LQC. Following then the main idea of QRLG, we try to repeat this procedure at the quantum level. As a result, we formulate a cosmological simplification of the LQG, without the spatial diffeomorphism and SU$(2)$ symmetries. This analog of the anisotropic extension of LQC will be tested in the forthcoming articles, verifying quantum connection between this model and LQG.

In particular, the paper is organized as follows. In Sec.~\ref{II}, we introduce lattice regularization in LQG, and provide a regularized expression for the geometric operators of the theory. Sec.~\ref{III} contains a general introduction to the reduction of canonical gravity. In Sec.~\ref{IV}, we reconsider the quantum reduction map derived in QRLG, and shed light on its inconsistencies and miscalculations. We also propose alternative reduction patterns, to avoid shortcomings, which are presented in Sec.~\ref{V}, demonstrating how LQC is linked with LQG. In Sec.~\ref{VI} we comment on the kinematical properties of a suitable quantum reduction of the general theory. Finally, in Sec.~\ref{VII} we present conclusions and outlooks on future investigations to be carried out.
\\

Through the paper, the metric signature is specified by $(-,+,+,+)$. The fundamental constant of LQG (representing the quantum of action analogously to $\hbar$ in quantum mechanics) reads $\kbar=\frac{1}{2}\gamma\hbar\kappa=8\pi\gamma l_P^2$, where $\gamma$ and $l_P$ are the Immirzi parameter and the Planck length respectively, and for simplicity we set the speed of light to $c=1$. The metric tensor, $g_{\mu\nu}=e^{\alpha}_{\mu}e^{\alpha}_{\nu}\eta_{\alpha\beta}$, can be cast in terms of $e^{\alpha}_{\mu}$, the co-vierbein fields, and the flat Minkowski metric reads $\eta_{\alpha\beta}$. The spatial metric tensor is denoted with $q_{ab}=e^i_a e^j_b\delta_{ij}$, where $e^i_a$ and $e_i^a=e^j_bq^{ab}\delta_{ij}$ denote co-dreibeins and dreibeins, respectively. A regularization via a cuboidal graph structure with the three directions of the orientation of links can be chosen, and adapted to the fiducial metric $^{0\!}q_{ab}$. Analogously, the constant orthonormal triad $^{0\!}e_i^a$ and co-triad $^{0\!}e^i_a$ are defined\footnote{The meaning of this structure should be clear. When we define the link between the volume of the fiducial cell $\mathbf{V}_{\!0}:=l_0^1l_0^2l_0^3$ and the physical volume $\mathbf{V}$ via the scale factor, we find that $a^3=\mathbf{V}/\mathbf{V}_{\!0}$. It is worth mentioning that considering the Bianchi I universe, the scale factor is defined as $a:=(a_1a_2a_3)^{{1}/{3}}$.}. Lowercase Latin indices $a,b,...=1,2,3$ label coordinate on each Cauchy hypersurface constructed by ADM decomposition \cite{Arnowitt:1962hi}, while $i,j,..=1,2,3$ are $\mathfrak{su}(2)$ internal indices and $\delta_{ij}$ stands for the Kronecker delta. Generators of $\mathfrak{su}(2)$ are defined as $\tau^i=-\frac{\i}{2}\sigma^i$, where $\sigma^i$ are Pauli matrices (see Appendix \ref{A}). Indices written in the bracket $(\,)$ are not summed, while for every other repeated pair the Einstein convention is applied.


\section{Regularization and operators}\label{II}

\noindent
In this section we introduce lattice regularization in LQG, and provide a regularized expression for the geometric operators of the theory. We start from classical general relativity, minimally coupled to the Standard Model of particle physics, and cast the theory within the Hamiltonian ADM formalism \cite{Arnowitt:1962hi}, in terms of the real Ashtekar-Barbero gravitational variables \cite{Ashtekar:1986yd}. Thus we consider the decomposition of the line element
\begin{align}\label{line_element}
ds^2:=g_{\mu\nu}dx^{\mu}dx^{\nu}
=(N^aN_a-N^2)dt^2+2N_adtdx^a+q_{ab}dx^adx^b\,,
\end{align}
where $N$ is the lapse function and $N^a$ the shift vector. We then look into the kinematical state space of the theory, discussing some subtleties of the reduction procedure.

\subsection{Classical theory}\label{II.1}

\noindent
We define the Einstein-Hilbert action with the cosmological constant term, minimally coupled to the free fields of the Standard Model,
\begin{align}\label{action_with_matter}
S:=S^{(g)}+S^{(\Lambda)}+\!\int_M\!\!\!\!d^{4}x\sqrt{-g}\,\mathcal{L}^{(\text{matter})},
\end{align}
where $\mathcal{L}^{(\text{matter})}$ encodes the Yang-Mills field, the scalar field, and the Dirac field.

In this section, we focus on the first two terms of $S$. The starting point in the construction of the Hamiltonian constraint operator (HCO) in LQG would be the Einstein-Hilbert action, which reproduces the classical equation of motion,
\begin{align}\label{action_EH}
S^{(g)}+S^{(\Lambda)}:=\frac{1}{\kappa}
\int_M\!\!\!\!d^{4}x\sqrt{-g}R-\frac{2\Lambda}{\kappa}\int_M\!\!\!\!d^{4}x\sqrt{-g},
\end{align}
where $R$ is the Ricci scalar. For completeness, we kept the cosmological constant $\Lambda$ in the action. Here $g$ stands for the determinant of the metric tensor $g_{\mu\nu}$, and the gravitational coupling constant reads $\kappa=16\pi G$.

The canonical quantization procedure in LQG is applied to the Hamiltonian obtained from action $S^{(g)}$, which is derived in the ADM formalism while using the Ashtekar variables. The latter are the Ashtekar-Barbero connection $A^i_a=\Gamma^i_a+\gamma K^i_a$ and the densitized dreibein $E_i^a=\sqrt{q}e_i^a$. Here, $\Gamma^i_a := \frac{1}{2}\epsilon^{ijk}\Gamma_{jka}= -\frac{1}{2}\epsilon^{ijk}e_k^b(\partial_ae^j_b-\Gamma^c_{ab}e^j_c)$ is the spin-connection and $\gamma K^i_a=\Gamma^i_{\ 0a}$ is the extrinsic curvature. The Ashtekar variables form a canonically conjugate pair, with a Poisson structure given by
\begin{align}
\label{Poisson_LQG}
\left\{A^i_a(t,{\bf x}), E_j^b(t,{\bf
y})\right\}=\frac{\gamma\kappa}{2}\delta_a^b\,\delta_j^i\,\delta^{(3)}({\bf x}-{\bf y})\,.
\end{align}
An important remark is that these variables are introduced by a canonical point transformation on the gravitational phase-space from the ADM canonical variables when the latter are written in the first-order form as $(K^i_a,E_i^a)$. From now on, for consistency of notation, we will use superscript ${}^{(A)}$ rather than ${}^{(g)}$, in order to denote objects describing gravitational degrees of freedom. Since we foliate spacetime and restrict our analysis to three-dimensional spatial hypersurfaces with metric tensor $q_{ab}$ on it, we reserve the term `metric' only to this object.

The Hamiltonian, which is obtained by the Legendre transform of \eqref{action_EH}, reads
\begin{align}
\label{A_total_Lambda}
H^{(A)}_T+H^{(\Lambda)}_T
=&\;\int_{\Sigma_t}\!\!\!\!d^3x\,\Big(
A^i_t\mathcal{G}_i^{(A)}
+N^a\mathcal{V}_a^{(A)}+N\big(\mathcal{H}^{(A)}+\mathcal{H}^{(\Lambda)}\big)\Big),
\intertext{where the three elements}
\label{A_Gauss}
G^{(A)}:=&\;\frac{1}{\gamma\kappa}\!\int_{\Sigma_t}\!\!\!\!d^3x\,A^i_t
D_aE^a_i,
\\
\label{A_vector}
V^{(A)}
:=&\;\frac{1}{\gamma\kappa}\!\int_{\Sigma_t}\!\!\!\!d^3x\,N^a
F^i_{ab}E^b_i,
\intertext{and}
\label{A_scalar}
H^{(A)}+H^{(\Lambda)}
:=&\;\frac{1}{\kappa}\!\int_{\Sigma_t}\!\!\!\!d^3x\,N
\bigg(
\frac{1}{\sqrt{q}}\big(F^i_{ab}-(\gamma^2+1)\epsilon_{ilm}K^l_aK^m_b\big)\epsilon^{ijk}E_j^aE_k^b
+2\Lambda\sqrt{q}
\bigg)
\end{align}
are called respectively the Gauss, the diffeomorphism (or vector) and the Hamiltonian (or scalar) constraints. These constraints impose respectively an internal SU$(2)$, a spatial diffeomorphism and a time reparametrization invariance. Hence the Hamiltonian constraint describes dynamics on the SU$(2)$ and spatial diffeomorphisms (or in short, diffeomorphisms) invariant subspace. Objects $A^i_t$, $N^a$ and $N$ are Lagrange multipliers. The quantity $F^i_{ab}$ denotes the curvature of the Ashtekar connection, while $D_a$ is a metric and dreibein compatible covariant derivative.

\subsection{Lattice regularization}\label{II.2}

\noindent
Lattice regularization in LQG is performed in two steps. In the first step, we begin from the imposition of the so-called `Thiemann trick', which goes as
\begin{align}
\label{E_trick}
\frac{1}{E_i^a}\big(\sqrt{|E|}\big)^{\!n}
=&\;\frac{2}{n}\frac{\delta\mathbf{V}^{n}}{\delta E_i^a}=\frac{4}{n\gamma\kappa}\big\{A^i_a,\mathbf{V}^{n}\big\}\,,
\\
K^i_a=&\;\frac{\delta\text{K}}{\delta E^a_i}=\frac{2}{\gamma\kappa}\big\{A^i_a,\text{K}\big\},
\end{align}
where $|E|=q$ is the absolute value of the determinant of $E^a_i$ and $\text{K}=\int\!d^3xK^i_aE^a_i$.

The second step is to regularize the spatial hypersurfaces \textit{via} a virtual granulation. It is realized by a construction of small solid objects --- grains, which fill all of the spacelike Cauchy hypersurface and intersect each other only in lower-dimensional submanifolds. This granulation of space is controlled by the parameter $\varepsilon$, where the limit $\varepsilon\to0$ corresponds to the granulating object of a trivial volume or, in other words, corresponds to taking the regulator to zero. This is done in a way similar to taking the decoupling limit in effective field theories --- decreasing the volume of the grains while at the same time increasing their number, in a way such that they always fill out the entire space. The standard choice for the shape of the solids is a tetrahedron. Consequently, the procedure is called a triangulation. The detailed description of this method can be found in \cite{Thiemann:1996aw,Thiemann:2007zz}. An alternative, much simpler choice for the shape of the solids is a cuboid --- albeit resulting in fixing some of the gauge freedom of the theory. This is the case of the `cubulation' procedure used in QRLG \cite{Alesci:2013xd}.

As a consequence of the granulation of the space, a regularization of the dynamical variables is introduced. After quantization, the effect of the regularization is to remove both the gravitational singularities (the initial singularity in a classical cosmology and the black hole singularity) and the UV-singularities of quantum matter fields \cite{Thiemann:2007zz}. Finally, there is an identification between the space and a graph $\Gamma$ that is created as a consequence of the granulation. This identification is realized by a duality: $\Gamma$ consists of links and nodes, hence in the dual graph we get respectively faces and volumes of the grains of $\Gamma^*$.

At the level of the canonical variables, the regularization is realized as follows. The Ashtekar connection $A^i_a$ is recovered from the holonomy $h_a(v):=h_{l^a}(v)$ (being a solution to the equation of a parallel transport of the connection) along the $l^a(v)$ link emanated from the $v$ node,
\begin{align}\label{line}
h_a(v)\!:=\mathcal{P}\exp\!\bigg(\int_{l^a}\!\!\!dsA^j_b(l(s))\tau^j\dot{l}^b(s)\bigg)\,.
\end{align}
Consequently, the curvature of the connection $F^i_{ab}(v)$ is turned into the holonomy around the loop $a\circlearrowleft b$ that starts from the initial point of link $l^a$, goes along this link and through the shortest polygon chain, it returns along link $l^b$ to the initial point. It is worth noting that the $a\circlearrowleft b$ loop is constructed, connecting paths along links $l^a$ and $(l^b)^{-1}$ intersected at the node $v$, with a path set by the arch $\alpha_{ab}$. Moreover, we assume to fix the orientation of this path according to the orientation of the given loop of links.

This regularization is realized via the relations
\begin{align}
\label{connection_to_holonomy}
h_p^{-1}\big\{h_p,{\bf V}\big\}=&\;\varepsilon\big\{A_a,\mathbf{V}\big\}\text{P}^a_p+\mathcal{O}(\varepsilon^2)\,,
\\
\label{curvature_to_holonomy}
2\epsilon^{pqr}h_{q\circlearrowleft r}=&\;\epsilon^{pqr}\big(h_{q\circlearrowleft r}-h_{q\circlearrowleft r}^{-1}\big)
=\epsilon^{ab(c)}\varepsilon^2F_{ab}\,\text{P}^p_c+\mathcal{O}(\varepsilon^4)\,,
\end{align}
where $F_{ab} = F^j_{ab}\tau_j$, $A_a = A_a^i\tau_i$ and $p,q,r...$ label directions of links of $\Gamma$, while $\text{P}^a_p$ is the projector onto these directions. As a result, we obtain all the gravitational dynamical variables written in terms of $h_p$, $h_{q\circlearrowleft r}$, $\mathbf{V}$ and (in the case of the gravitational contribution to the Hamiltonian constraint) $\text{K}$. Namely, the scalar constraint density reads,
\begin{align}
\begin{split}\label{g_lattice_scalar}
H^{(A)}=&\;\frac{1}{\kappa}\lim_{\varepsilon\to0}\frac{1}{\varepsilon^3}\!\int\!\!d^3x\,N\epsilon^{pqr}\bigg(
\frac{2^3}{\gamma\kappa}\,\text{tr}\Big(h_{p\circlearrowleft q}\,h_r^{-1}\big\{\mathbf{V},h_r\big\}\Big)
\\
&\;-\frac{2^5(\gamma^2+1)}{\gamma^3\kappa^3}\,\text{tr}\Big(
h_p^{-1}\big\{\text{K},h_p\big\}\,h_q^{-1}\big\{\text{K},h_q\big\}\,h_r^{-1}\big\{\mathbf{V},h_r\big\}\Big)\bigg).
\end{split}
\end{align}
Having derived the lattice-regulated scalar constraint, the quantization method is straightforward and can be implemented \textit{via} the Dirac procedure \cite{Dirac:1950pj}. We turn Poisson brackets into commutators multiplied by $1/\i\hbar$ and change the dynamical variables into operators. The latter ones are the holonomy and flux of the densitized dreibein operator, as well as the geometrical operators, \textit{i.e.} the volume, the area and the length operators. It is worth mentioning that the holonomy and flux are the canonical pair, while all the geometrical operators are constructed as smeared appropriate combinations of the densitized dreibein operators.

The kinematical Hilbert space of LQG (and also in QRLG) is a direct sum of cylindrical functions of (possibly reduced, \textit{i.e.} diagonal) connections along the links of the graph $\Gamma$ (the general graph or cuboidal one ${}^R\Gamma$ in the reduced case). In the case of LQG, the kinematical Hilbert space is equipped with an inner product defined as an integral over cylindrical functions with an SU$(2)$-invariant Haar measure.

\subsection{DeWitt coordinate representation}\label{II.3}

\noindent
The Alesci-Cianfrani model is based on an appropriate projection of the SU$(2)$ group elements to the three U$(1)_{p}$ subgroups defined along the directions of the $t^p:=\!\vec{\,\rho}^p\cdot\vec{t}$ basis vectors, constructed as rotations of $t^3=u^3$ (being one of the Lie algebra generators) into the unit vector $\!\vec{\,\rho}^p$ that projects $t^i$ onto $t^3$ and $\tau^i$ onto $\tau^3$, respectively --- see Appendices \ref{A} and \ref{B}. This projection is based on the Livine-Speziale SU$(2)$ coherent states \cite{Livine:2007vk}, which are defined along $t^p=\text{P}^p_it^i$ and minimize uncertainty of the gravitational momentum operator (defined in \eqref{DeWitt_E}).

QRLG has been constructed as a cosmological model with the general relativistic diffeomorphism invariance broken down to the Bianchi I symmetry, supposedly in a more rigorously defined way than in the case of LQC. The latter one, already before the quantization, replaces the Ashtekar connection in the definition of holonomy \eqref{line} with a diagonal variable defined as $A^{(i)}_a|_{_\text{LQC}}:=\bar{c}_{(i)}{}^{0\!}e^{(i)}_a\big/l_0^{(i)}$ \cite{Ashtekar:2003hd}, with $\varepsilon\,l_0^{(i)}$ being the length of the ${}^{0\!}e_{(i)}^a$ side of the fiducial elementary cell, while $\bar{c}_{(i)}$ is a constant. As a result, the real SU$(2)$ holonomy becomes replaced with a complex one, invariant under U$(1)$ transformations. Considering a link $l^{(i)}$ of the length $\varepsilon l_0^{(i)}$, one finds the explicit form of the LQC holonomy,
\begin{align}
\label{LQC_holonmy}
h_{(i)}|_{_\text{LQC}}
=\bigg(\frac{1}{2}-\i\tau^{(i)}\bigg)\exp\!\bigg(\frac{\i}{2}\varepsilon\,\bar{c}_{(i)}\bigg)
+\bigg(\frac{1}{2}+\i\tau^{(i)}\bigg)\exp\!\bigg(\!-\frac{\i}{2}\varepsilon\,\bar{c}_{(i)}\bigg)
=\exp\big(\varepsilon\,\bar{c}_{(i)}\tau^{(i)}\big)\,,
\end{align}
where the object $\exp\!\big(\!\pm\!\frac{\i}{2}\varepsilon\,\bar{c}_{(i)}\big)$ is the complex U$(1)$ holonomy. It is worth mentioning that an explicit form of the factor $\varepsilon$ distinguishes between so called $\mu_0$-scheme \cite{Ashtekar:2003hd,Bojowald:2008zzb} or $\bar{\mu}$-scheme \cite{Ashtekar:2006uz,Ashtekar:2006wn}.

Notice that to recover a similar appearance of the holonomy as in LQC, Alesci and Cianfrani redefined the real LQG holonomy \eqref{line} to be the imaginary one $h_{l}=\exp\!\big(\!\pm\!\i\!\int_{l}A^j_as^j\dot{l}^a\big)$. Then to preserve the structure of the theory, \textit{i.e.} $h_{l}\in\text{SU}(2)$, they replaced the generators of the $\mathfrak{su}(2)$ representation with the self-adjoint operators $s^i:=\sigma^i/2$ c (see Appendix \ref{A}). To avoid confusion, we are going to keep the standard notation, and show that there is no difference between $\exp\!\big(\!\pm\!\frac{\i}{2}\varepsilon\bar{c}_{(i)}\big)$ and the reduced real LQG holonomy in \eqref{line}. In both cases, the action of the densitized dreibein operator leads to a real eigenvalue.

In LQG, QRLG and LQC we use the DeWitt-like representation \cite{DeWitt:1967yk} of the Ashtekar variables,
\begin{align}
\label{DeWitt_A}
A^i_a\to&\ \hat{A}^i_a|\ldots\rangle=A^i_a|\ldots\rangle\,,
\\
\label{DeWitt_E}
E_i^a\to&\ \hat{E}_i^a|\ldots\rangle=-\i\kbar\frac{\delta}{\delta A^i_a}|\ldots\rangle\,.
\end{align}
Here, $|\ldots\rangle$ denotes a standard basis vector in LQG or QRLG, which is defined in Sec.~\ref{II.4}. Notice that the operators in \eqref{DeWitt_A} and \eqref{DeWitt_E} do not correspond geometrically to their classical equivalents, since the Ashtekar connection has dimension of a length$^{-1}$, while the eigenvalue of the densitized dreibein operator has dimension of a $\kbar\,\times\,$length, and thus length$^{3}$.

The proper rescaling has been suggested by LQC \cite{Ashtekar:2003hd}, and later adapted to QRLG extending Bianchi I metric to an inhomogeneous model \cite{Alesci:2013xd}. It defines a pair of canonical variables $(c_{(i)},p^{(j)})$, which in the case of LQC are spatially constant, $c_{(i)}\to\bar{c}_{(i)}$, $p^{(j)}\to\bar{p}^{(j)}$. The map from the Ashtekar variables to the reduced ones, $(A^i_a\to{}^R\!A^{i}_a,E_i^a\to{}^R\!E_{i}^a)$, is defined as follows,
\begin{align}
\label{QRLG_A}
{}^R\!A^{i}_a(t,{\bf x}):=&\;\frac{1}{l_0^{(i)}\!}c_{(i)}\big(t,{\bf x}\big){}^{0\!}e^{i}_a\,,
\\
\label{QRLG_E}
{}^R\!E_{i}^a(t,{\bf x}):=&\;\frac{l_0^{(i)}}{{\bf V}_{\!0}}p^{(i)}\big(t,{\bf x}\big)\sqrt{^{0\!}q}\,{}^{0\!}e_{i}^a\,,
\end{align}
where ${\bf V}_{\!0}:=l_0^1\,l_0^2\,l_0^3$. The canonical Poisson relation for the reduced variables is summarized in Appendix \ref{C}. It is worth mentioning that in QRLG, the inhomogeneous extension of Bianchi I appears to be controllable imposing the diffeomorphism constraint on the partially gauge-fixed and partially reduced states space (see Sec.~\ref{IV}). This leads to the reduced and constrained variables,
\begin{align}
\label{R-C_A}
{}^{R\text{-}C}\!\!A^{i}_a(t,{\bf x}):=&\;\frac{1}{l_0^{(i)}\!}c_{(i)}\big(t,x^{(i)}\big){}^{0\!}e^{i}_a\,,
\\
\label{R-C_E}
{}^{R\text{-}C}\!E_{i}^a(t,{\bf x}):=&\;\frac{l_0^{(i)}}{{\bf V}_{\!0}}p^{(i)}\big(t,x^{(i)}\big)\sqrt{^{0\!}q}\,{}^{0\!}e_{i}^a\,.
\end{align}
However, as pointed in Sec.~\ref{IV}, the inhomogeneous extension is a source of over-constraining for the theory.

Now we are ready to prove that the eigenvalue of the $\hat{p}^i$ operator acting on the state based on a reduced holonomy is real. For simplicity, we assume that the holonomy is oriented along the third internal direction, \textit{i.e.} along the link $l^3$. Following Alesci and Cianfrani \cite{Alesci:2013xd,Alesci:2013xya,Alesci:2014uha,Alesci:2016gub,Alesci:2017kzc}, we impose the DeWitt representation \eqref{DeWitt_E} --- this is not completely rigorous (see Sec.~\ref{IV}-\ref{VI}), but is sufficient for our purposes --- getting
\begin{align}
\label{reality}
\hat{p}^i\Big|e^{\varepsilon c_{3}\tau^{3}}\Big>_{\!\!_R}
=-\i\kbar\frac{\delta}{\delta c_i}\Big|e^{\varepsilon c_{3}\tau^{3}}\Big>_{\!\!_R}
=-\i\kbar\:\!\varepsilon\:\!\delta_3^i\Big<\tau^{3}\Big|e^{\varepsilon c_{3}\tau^{3}}\Big>_{\!\!_R}
=-m\kbar\:\!\varepsilon\:\!\delta_3^i\Big|e^{\varepsilon c_{3}\tau^{3}}\Big>_{\!\!_R}\,.
\end{align}
Notice also that in the last step we used the $-im$ eigenvalue of the $\tau^3$ generator in the spherical basis (see \eqref{upsilon123}). The result in \eqref{reality} coincides (up to the sign, being dependent on the orientation of links, and a result of a convention in a definition of the Lie algebra generators) with all the articles considering QRLG.

\subsection{State space: SU$(2)$-coherent spin-network}\label{II.4}

\noindent
Another feature of the Alesci-Cianfrani construction, apart form the rotation onto the $t^p$ directions, is the modification of intertwiners. They are the components of the kinematical Gauss-invariant Hilbert space of LQG, introduced as a result of the implementation of the Gauss constraint \eqref{A_Gauss} at the quantum level. The intertwiners, in the definition of spin-network states (the basis states of the theory), connect SU$(2)$ irreducible representations attached to the links of the $\Gamma$ graph. Consequently, intertwiners are thought to be attached to the nodes of $\Gamma$, and act as projectors enforcing the SU$(2)$ gauge invariance via the group averaging. Within the case of the spin-network states rotated onto three orthogonal directions, unique intertwiners are represented by the Clebsch-Gordan coefficients, or equivalently by the Wigner 3-$j$ symbols. It is worth noting that intertwiners do not appear in LQC. Therefore QRLG, as long as it is formulated by the projection on the Gauss-invariant Hilbert space of LQG, appears to retain a more adherent structure to the original theory than LQC.

The spin-network states, defined as $\langle h|\Gamma;j^l,i_v\rangle$, are supported on the graphs $\Gamma$, labelled by the spins $j^l$ (encoding $\mathfrak{su}(2)$ irreducible representations of the holonomies 
attached to each links $l$ of $\Gamma$) and the intertwiners $i_v$ (implementing SU$(2)$ invariance at each node $v$ of $\Gamma$). In QRLG the basis-like states are $_{_R}\!\langle\bar{h}|\Gamma;j^p,i_v\rangle\!_{_R}$ and involve Wigner matrices, which are rotated to the $t^p$ directions and projected on the coherent Livine-Speziale states \cite{Livine:2007vk}, with maximal spin number,
\begin{align}\label{spin_number}
j^p\ \to\ \bar{j}^p:=j^p_{\text{max}}\,.
\end{align}
Moreover, the secondary spin quantum number (called also the magnetic number), taking values $m^p=-j^p,-j^p+1,...,j^p$ and being the eigenvalue of the $s^3$ internal angular momentum generator (as in particle physics) in the spherical basis (this notation is explained in Appendix \ref{A}), is fixed to
\begin{align}\label{magnetic_number}
m^p\ \to\ \bar{m}^p:=\pm\bar{j}^p\,.
\end{align}

The reduced spin-network space is constructed as a modified space of solutions of the constraints \eqref{A_Gauss} and \eqref{A_vector} and will be denoted as $^{R\!}\mathcal{H}_{\Gamma,v}^{(gr)}$. In the case of the Alesci-Cianfrani model, the SU$(2)$ invariance is replaced with the three U$(1)$ symmetries along the directions of the links of ${}^R\Gamma$, while the diffeomorphism constraint is restricted to the implementation of an invariance under spatial diffeomorphisms, which do not generate any off-diagonal components. The former restriction is an internal gauge fixing realized by the projection on the coherent states. The latter one can be interpreted as an external gauge fixing of the geometry, which restricts a generic $\Gamma$ graph to the cuboidal one, ${}^R\Gamma$. A precise construction of the Hilbert space of the full theory, LQG, $\mathcal{H}_{kin}^{(gr)}$, can be found \textit{\textit{e.g.}} in \cite{Thiemann:2007zz}, while that for QRLG is given in \cite{Alesci:2013xd,Alesci:2013xya}. Details of the mechanism and consequences of rotational transformations imposed on the LQG spin-network are discussed in the Sec.~\ref{IV.1}.

Finally, the problem of solving the Hamiltonian constraint at the quantum level recasts as the problem of finding solutions of the action $\hat{H}|\Gamma;J,I\rangle$ --- in the reduced case, the action $^{R\!}\hat{H}|\Gamma;J,I\rangle\!_{_R}$. Here, $J\ni m^l$ and $I\ni i_v$ are the set of spin numbers and the set of intertwiners, attached respectively to all links and all nodes of a given graph. For simplicity, we omitted labeling with `$^{R}$' the quantities $\Gamma$, $J$ and $I$ inside the $|...\rangle\!_{_R}$ `kets' describing reduced states.

As we already mentioned, the projection of the Wigner matrices on the coherent states, simultaneously projects SU$(2)$ intertwiners. As a consequence, states become decomposed as follows,
\begin{align}
\label{unnormalized}
{}_{_R}\!\langle\bar{h}|\Gamma;J,I\rangle\!_{_R}
:=
\prod_{v\in\Gamma}\big<j^l\!,i_v\big|\bar{m}^p\!,t^p\big>\prod_{l^p\in\Gamma}\!^{p}\!D^{|\bar{m}|^p}_{\!\bar{m}^p\,\bar{m}^p}(\bar{h}_p)\,,
\end{align}
where $\langle j^l\!,i_v|\bar{m}^p\!,t^p\rangle$ are the reduced U$(1)$ intertwiners, while $^{p}\!D^{|\bar{m}|^p}_{\!\bar{m}^p\,\bar{m}^p}(\bar{h}_p)$ is the Wigner $D$-matrix, with a fixed irreducible representation $j^p\to\bar{j}^p=|\bar{m}|^p$ attached to the $l^i$ link.

It is important to notice that the basis-like states are not orthonormal within the scalar product given by the expression
\begin{align}\label{scalar_product}
{\phantom{\big<}}_{_R\!\!}\big<\Gamma;m^{l^p}\!,i_v\big|\Gamma'\!,m^{{l'}^q}\!\!,i_v'\big>_{\!\!_R}
=\delta_{\Gamma,\Gamma'}\prod_{v\in\Gamma}\prod_{l\in\Gamma}\delta_{j^{l^p}\!\!,\,j^{{l'}^q}}
\big<\bar{m}^{l^p}\!,t^p\big|j^{l^p}\!,i_v\big>\big<j^{{l'}^q}\!\!,i_v'\big|m^{{l'}^q}\!\!,t^q\big>\,.
\end{align}
The term $\big<\bar{m}^{l^p}\!,t^p\big|j^{l^p}\!,i_v\big>\big<j^{{l'}^q}\!\!,i_v'\big|m^{{l'}^q}\!\!,t^q\big>$ represents a product of U$(1)$ phases. It is also worth mentioning that by definition any Hilbert space is complete, \textit{i.e.} it has an orthonormal basis. Therefore we suggest to impose by hand the following normalization
\begin{align}
\label{basis_vector}
|\Gamma;J\rangle\!_{_R}:=
\prod_{v\in\Gamma}\prod_{l\in\Gamma}\big(\big<j^l\!,i_v\big|\bar{m}^p\!,t^p\big>\big)^{\!-1}|\Gamma;J,I\rangle\!_{_R}\,,
\end{align}
to drop the phase dependence form the non-orthonormal states obtained by the reduction procedure. As a result, the normalized state space of QRLG, namely $^{R\!}\mathcal{H}_{\Gamma,v}^{(gr)}$, gets rid of its dependence on intertwiners placed at the nodes $v$. Therefore this can be understood as a Hilbert space, becoming $^{R\!}\mathcal{H}_{\Gamma}^{(gr)}=\otimes_{l^p\in \Gamma} \mathcal{H}_p$, with $\mathcal{H}_p$ denoting the U$(1)_{p}$ Hilbert space associated to each orthogonal direction. In other words, we simplify the structure of the sum over intertwiners into a contraction of Kronecker delta functions oriented along the link-directions.

Let us present one more argument why the state provided by expression \eqref{basis_vector} can be considered as the spin-network of QRLG. From the point of view of the Dirac program of canonical quantization of constrained systems \cite{Dirac:1950pj}, one should impose constraints one by one, to recover the physical phase-space. Notice that for the $\text{U}^3(1)$ symmetry, the vector constraint vanishes identically\footnote{Precisely speaking, reduced diffeomorphisms map directions of links into themselves. Hence for a diagonal form of the dreibein, the directions restrict the lattice to be cuboidal. Therefore fixing holonomies to the ones of the diagonal connections, which are attached to the cuboidal lattice, we neglect the vector constraint.}. The Gauss constraint becomes reduced to abelian transformations of Lie group along three orthogonal directions, $\bar{h}_i\to\bar{h}_i'=g_{(i)}\bar{h}_ig_{(i)}^{-1}=\bar{h}_i$, where $\bar{h}_i,g_{(i)}\in\text{U}(1)$. Hence there is no reason to introduce the construction of intertwiners (which is necessary in LQG formulated in terms of SU$(2)$ group elements). Then from the geometrical perspective of the SU$(2)$ to U$(1)$ reduced theory, these transformations are simply the phase transformations, with generators being $\mathds{C}$ numbers. This allows us to perform the normalization as defined in \eqref{basis_vector}. Another way to reproduce this result is moving the reduced intertwiner $\langle j^l\!,i_v|\bar{m}^p\!,t^p\rangle$ in \eqref{unnormalized} to the right hand-side of the expression, and then rescaling the U$(1)$ holonomy. For consistency, let us assume to move all the intertwiners in the unnormalized space into holonomies attached to the links emanated towards positive orientation. Then, since for a given node-link pair, the intertwiners are fixed spin $\bar{m}^i$-dependent functions, we simply rescale appropriately $\varepsilon$ in $\big|e^{\varepsilon\tilde{c}_{i}\tau^{(i)}}\big>_{\!\!_R}$.

Finally, let us discuss why the reduced intertwiners, which appear to be only a redundant complication, are still present after the SU$(2)$ to U$(1)$ reduction in the original formulation of QRLG \cite{Alesci:2013xd,Alesci:2013xya,Alesci:2014uha,Alesci:2016gub,Alesci:2017kzc}. These are a consequence of the reduction of a partially constrained kinematical Hilbert space of LQG. Notice that all the constraints, including the Gauss \eqref{A_Gauss}, the diffeomorphism \eqref{A_vector} and the Hamiltonian one \eqref{A_scalar}, are first class secondary constraints. They are independent, therefore after the quantization they should be imposed on the spin-network in any order, but necessarily during the same step, without any manipulations on the structure of the Hilbert space after implementation of only one of the constraints. As a result we would obtain a physical Hilbert space. Only by convenience --- to simplify calculations --- we first impose the Gauss constraint, then the diffeomorphism one and finally the Hamiltonian constraint. Hence for a consistency, the reduction procedure should be performed either on the kinematical, or on the physical Hilbert space and not on the gauge invariant kinematical Hilbert space (after imposition of only the Gauss constraint). The former choice does not generate the reduced intertwiners, because the kinematical Hilbert space is the space of cylindrical functions over the $\Gamma$ graph, equipped with the Ashtekar-Lewandowski measure \cite{Ashtekar:1994mh,Ashtekar:1994wa}, without yet introduced the SU$(2)$ intertwiners. The latter choice, \textit{i.e.} the reduction of the physical Hilbert space --- up to the present stage of the development of LQG --- is impossible to achieve. The Hamiltonian of LQG is so complicated that the full structure of the physical Hilbert space remains unknown. This argumentation is developed in further analyses, contained in Sec.~\ref{IV}, while in Sec.~\ref{V} where we present the proper order of implementation of the constraints leading to a well defined and simplified theory.

\subsection{Gravitational field operators in LQG}\label{II.5}

\noindent
Let us now discuss the generic model of LQG. While taking into account the cosmological constant's sector, the whole difficulty in finding a solution to the equation $H^{(\Lambda)}|\Gamma;j^l,i_v\rangle$ becomes the derivation of the action of the volume operator,
\begin{align}\label{volume}
\hat{\mathbf{V}}|\Gamma;J,I\rangle\,.
\end{align}
The gravitational Hamiltonian $H^{(A)}$ produces two classes of equations for the eigenvalues of the $\mathfrak{su}(2)$ traces of the operators in \eqref{g_lattice_scalar}.  As usual in the standard literature, we are going to call the first one the Euclidean term,
\begin{align}\label{loop}
\,\text{tr}\bigg(\hat{h}_{p\circlearrowleft q}\,\hat{h}_r^{-1}\Big[\hat{\mathbf{V}}_{\!v},\hat{h}_r\Big]\bigg)|\Gamma;J,I\rangle \,.
\end{align}
The second, being the most complicated object, has been named the Lorentzian term and it is given by the formula
\begin{align}\label{curvature}
\,\text{tr}\bigg(\hat{h}_p^{-1}\Big[\hat{\text{K}}_v,\hat{h}_p\Big]\hat{h}_q^{-1}\Big[\hat{\text{K}}_v,\hat{h}_q\Big]
\,\hat{h}_r^{-1}\Big[\hat{\mathbf{V}}_{\!v},\hat{h}_r\Big]\bigg)|\Gamma;J,I\rangle\,.
\end{align}

It is worth noting that equation \eqref{volume} is solvable for simple configurations of states. However a problem, which arises is the fact that there is an ambiguity in the choice of the definition of the volume operator \cite{Lewandowski:1996gk,Flori:2008nw}. Besides that, in order to derive actions of the complete set of all the Standard Model matter fields, we need rather some powers of $\hat{\mathbf{V}}_{\!v}$ \cite{Thiemann:1997rt}. Hence, instead of focusing only on the cosmological constant sector described by formula \eqref{volume}, we need to solve the following action,
\begin{align}\label{volume_n}
\big(\hat{\mathbf{V}}_{\!v}\big)^{\!n}|\Gamma;J,I\rangle\,,
\end{align}
$n$ being a positive rational number.

In the case of equation \eqref{loop}, the solutions for standard LQG have been found only for single-node states of a particular valency\footnote{See \textit{e.g.} \cite{Alesci:2013kpa} (for trivalent nodes) or \cite{Alesci:2011ia} (for tetravalent nodes).} and for coherent complexifier states\footnote{The Dapor-Liegener model called Cosmological Complexifier-Coherent Loop Quantum Gravity has been first proposed in \cite{Dapor:2017rwv}. Its initial construction has been explained in details in \cite{Dapor:2017gdk}. Nonetheless, further investigations are required in order to show how the formalism of LQC is recovered.}. However, in the case of the reduced graph, a general solution exists \cite{Alesci:2014uha}. This latter takes a simpler form upon inclusion of the corrections from the reduction procedure discussed in section \eqref{IV}.

Derivation of the Lorentzian term in equation \eqref{curvature} is even more demanding. As in the case of the Euclidean contribution to the Hamiltonian constraint, the result for a general case with a big number of nodes of different valency is rather impossible to be achieved. In the reduced model at the classical level this term does not appear any more. This is a consequence of the diagonalization of the spatial metric tensor (which is a correct assumption if one considers only a leading order term in the semiclassical analysis of QRLG). However, a precise approach to quantization of the Hamiltonian constraint has to be applied to the Lorentzian term as well. We expect that the next to the leading order corrections to the matrix element, expanded around the classical configuration, are of the same order of significance as the corrections from the expansion of the Euclidean term. Moreover, the corrections from the expansion of the Lorentzian term in the framework of QRLG could be different with respect to the ones obtained by Alesci and Cianfrani in \cite{Alesci:2014uha}, as it happens in the case of different approaches to LQC\footnote{The next to the leading order corrections arising from the Euclidean and Lorentzian terms differ among each other. Up to our knowledge, there exist at least three paths to recover the original formulation of LQC (considering only the Euclidean term), including the frameworks proposed in \cite{Yang:2009fp} and \cite{Dapor:2017rwv}.}.


\section{Cosmological reduction of canonical gravity}\label{III}

\noindent
Before discussing the cosmological reduction of LQG, let us analyze from a general perspective what are the possible approaches to this issue. The problem is clear and already solved at the classical level. The cosmological reduction imposes the vanishing of the Gauss and diffeomorphism constraints, thus it leads to the quantization of the classically gauge-invariant functions already restricted to the equivalence classes of systems linked by gauge transformations. Its most complete description in the homogeneous and isotropic case is the formulation of LQC presented in \cite{Ashtekar:2011ni}. The anisotropic extension of this theory is given in \cite{Ashtekar:2009vc}. In these models, the only remaining operator equation is the scalar constraint that acts on states, which must be selected to satisfy the time reparametrization invariance. However, this model has not been verified from the perspective of the formal reduction of the phase space, imposing canonical gauge conditions. We fill this gap, describing an appropriate procedure in Sec.~\ref{V.1}.

It should be also clear that the most complete approach at the quantum level would be to impose the cosmological symmetry on the general solution of LQG. However, since this solution does not exist, we should try to reduce --- by a well defined and controlled method --- the system of not (totally) solved quantum constraints. When discussing the general procedure we follow the Henneaux-Teitelboim recipe \cite{Henneaux:1992ig}.

Before classifying the possible approaches, let us formulate few universal statements. Any first class constraint (there are three of them in the case of the canonical gravity) `hits twice'. Precisely speaking, given the Hamiltonian formulation of any theory,  each constraint must be solved as an independent operator equation (gauge condition) and each gauge symmetry must be realized by states and operators (they both have to be transformation invariant). In other words, an operator that modifies the states cannot send these states outside the gauge invariant Hilbert space. The formal reduction of the phase space requires the introduction of gauge conditions. These conditions are constraints introduced \textit{ad hoc}, constituting a second class system. It is worth mentioning that the number of the gauge conditions must be equal or greater than the number of the original first class constraints. Moreover, in order to avoid the Gibov obstruction --- see Appendix \hyperref[D.1]{D.1} --- the gauge conditions have to be implemented globally. One should then replace Poisson brackets with Dirac brackets. The procedure of quantization of the secondary constraints and Dirac brackets is however not uniquely solved, thus we are going to discuss the methods to avoid this problem.

The easiest, but not formal way to get around this issue is to remove the symmetry and associated constraints `manually' at the classical level --- discuss this method in \hyperref[D.1]{D.1}. As a result, one consider a different, simpler theory from the beginning and only this theory is quantized, without a connection with the general model that has been simplified. A good example of this strategy are LQC (when comparing with LQG) and time gauge in the tetrad formalism of the $3+1$ gravity (when compared with the Palatini formulation). Notice that a formal way to get around the problem is a redefinition of the canonical quantization --- described in \hyperref[D.2]{D.2}. We show how this straightforwardly solves the difficulties of the gauge conditions that are recalled in Sec.~\ref{V.1}.

A more involving method is a reduced phase space quantization, explained in \hyperref[D.3]{D.3}. In Sec.~\ref{V.1} we show how imposing appropriate gauge conditions, Gauss and spatial diffeomorphism invariance are frozen, leading to the anisotropic formulation of LQC. Up to our knowledge this is the first proof that the gravitational sector of LQC is a gauge-reduced equivalent version of LQG.

Finally, it is interesting to check whether imposing the reduced phase space technique at the quantum level in terms of operators, one would obtain the same result\footnote{We remark that there is no procedure that uniquely specify how to implement constraints at the quantum level.}. In Sec.~\ref{V.2} and Sec.~\ref{V.3} we demonstrate that solving gauge conditions as operators, one finds isomorphic Hilbert space to the one defined for the classically reduced phase space. What remains unknown is if the original constraint operators and the corresponding states, not solved during fixation of gauge symmetries, are quantum-reduced to the same eigenequations. For instance, while considering the reduction of LQG at the quantum level, the question of whether the eigenequation containing the Lorentzian contribution to the scalar constraint operator is based on the same reduced states as the ones contributing to the Euclidean eigenequation, remains open after our analysis. From the structure of the reduced Hilbert space, we anticipate that the states in both cases are indeed the same. The corresponding matrix elements, however, may be different and until this issue will not be explicitly verified, we are not able to formulate even a heuristic guess what will be the answer.
\\

Before starting our analysis, let us check whether some of these issues are already solved. The earlier attempt to investigate these problems was QRLG. However, we found this model to be incorrectly formulated, what we demonstrate in the next section.


\section{Quantum reduction of spin-network}\label{IV}

\noindent
QRLG has been constructed as an alternative to LQC. It was thought to retain a definite advantage with respect to the latter theory, since it was believed to provide a precisely defined reduction procedure of LQG at the quantum level. It comes together with significant simplifications with respect to the full theory. As was already mentioned, formulas (\ref{loop}), (\ref{curvature}) and (\ref{volume_n}), which altogether appear to be un-tractable in the general case of LQG, have analytical solutions in QRLG.

Let us begin with the volume operator. The regularized action of this operator $\eqref{volume_n}$ in full LQG has complicated structure \cite{Thiemann:2007zz,Ashtekar:1997fb}. Neglecting ineffectiveness (from the point of view of applications) of a direct regularization of fluxes, we can write the action of $\hat{\mathbf{V}}_{\!v}$ defined around a neighborhood centered at the $v$ node as
\begin{align}\label{volume_operator}
\big(\hat{\mathbf{V}}_{\!v}\big)^{\!n}\big|\Gamma;j^l,i_v\big>_{\!\!_R}
=\Bigg[\!\int\!\!d^3x
\bigg(\bigg|\frac{1}{3!}\epsilon^{ijk}\epsilon_{pqr}\hat{E}_i({\bf S}^p)\hat{E}_j({\bf S}^q)\hat{E}_k({\bf S}^r)\bigg|\bigg)^{\!\!\frac{1}{2}}
\Bigg]^{\!n}
\big|\Gamma;j^l,i_v\big>_{\!\!_R}\,,
\end{align}
where it has been assumed that the operator of a volume to a given power equals that power of the volume operator\footnote{This assumption is better legitimated in QRLG, where the volume operator is an eigenoperator of the reduced spin-network (see analogous comments in Sec. \ref{VI}).}. The irregularity and complication of a structure of the general graph directly prevents from getting solution to equation \eqref{volume_operator}. Since the same operator appears in other equations such as \eqref{loop} and \eqref{curvature}, and in the gravitational contributions to the HCO's of matter fields, it follows that these cannot be solved either. The situation is much simpler in the Alesci-Cianfrani model with the regular cuboidal, self-dual graph. 

The self-duality of ${}^R\Gamma$ should be understood in a geometrical way. The faces dual to the links of ${}^R\Gamma$ and the polyhedra dual to the nodes of this graph are respectively rectangles and cuboids. They are elements of the dual space. Then, the dual graph ${}^R\Gamma^*$ is constructed from the edges and vertices of the cuboids. The result is the ${}^R\Gamma^*$ graph, congruent to ${}^R\Gamma$. Moreover, identifying the edges and vertices with some lattice's links and nodes, respectively, leads to an analogous structure to ${}^{R\!}\mathcal{H}_{\Gamma}^{(gr)}$, where $^{R\!}\mathcal{H}_{\Gamma}^{(gr)}$ is the normalized Hilbert space of QRLG. Then one can choose some averaging procedure that translates $|\bar{m}|^i$ spin numbers attached to ${}^R\Gamma$ onto the ones along the links of ${}^R\Gamma^*$, emanated from the nodes shifted by a half link distance. As a result, one obtains a Hilbert space $^{R\!}\mathcal{H}_{\Gamma^*}^{(gr)}\cong{}^{R\!}\mathcal{H}_{\Gamma}^{(gr)}$. It is worth noting that a similar identification for the state space of QRLG including intertwiners, $^{R\!}\mathcal{H}_{\Gamma,v}^{(gr)}$, is generally not true --- except for the homogeneous case, in which this identification is natural. This identification is not correct due to the presence of intertwiners placed at nodes of ${}^R\Gamma^*$, which should not be related to the ones in ${}^R\Gamma$, but should provide a gauge invariance in $^{R\!}\mathcal{H}_{\Gamma^*\!,v}^{(gr)}$.

Another relevant feature of ${}^R\Gamma$ is that it is a fixed graph, conversely to the graph structure, which supports LQG. The latter one is the uncountable (almost direct) sum of disjoint graphs, hence it is non-separable. Thus it can represent continuous geometries, being embedded in a differential manifold. The former one decomposes into a direct product of three fixed graphs. Each one supports a family of states, which corresponds to a fixed one-dimensional geometry. Moreover, the reduction procedure restricts both the canonical pair as well as all the geometrical operators to the ones that preserve the structure of ${}^R\Gamma$. Therefore each space of cylindrical functions over a fixed one-dimensional lattice is a superselection sector with corresponding graph-preserving (also called non-graph-changing) operators. Expectation values of these operators (if well constructed\footnote{Notice that although a certain set of operators may not satisfy the definition of being the Dirac observables, their expectation values are classical quantities, and thus constitute the obserables. It is worth mentioning that in many formulations of LQG (see \textit{e.g.} \cite{Thiemann:1996aw}) and in the minisuperspace-based reduced variants of the theory (see \textit{e.g.} \cite{Ashtekar:2003hd}), the hermiticity of operators associated to classical quantities that are measurable, is not an \textit{a priori} included constructional condition --- see \textit{e.g.} the case of the Hamiltonian operator. Notice, however, that the main inventors of both LQG and LQC included the hermiticity as a constructional requirement for the canonical pair of operators and all the geometrical operators --- see \cite{Thiemann:1996aw,Ashtekar:1997fb,Thiemann:2007zz} and \cite{Ashtekar:2003hd,Ashtekar:2011ni}, respectively. Here, by the ``well constructed operator'', we understand their self-adjoint versions, easy obtainable by adding the adjoint operator to any non-Hermitian operator, for which exists its classical equivalent.}) are Dirac observables. Each of these sectors is equipped with the U$(1)$ Haar measure on the Bohr compactifications of the real line \cite{Alesci:2013xd}. It is also worth mentioning that in the homogeneous limit, this polymer-like structure simplifies into a collection of lines, equipped with the Lebesgue measures.

Notice that the kinematical Hilbert space of LQC is not separable by an analogous argument. The main difference is that the action of HCO connects different superselection sectors \cite{Ashtekar:2006uz}. Then HCO is modified to preserve these sectors, while the physical Hilbert space is constructed from the states on which this HCO acts\footnote{Another possible solution that tames this problem is called the integral Hilbert space method. In this model a separable Hilbert space for LQC is constructed in terms of an integral of superselection sectors equipped with a Lebesgue measure (see \cite{G.:2014lpa}).}.

\subsection{Reduction procedure I: projected space in QRLG}\label{IV.1}

\noindent
The first complete description of the reduction of LQG to QRLG was shown in \cite{Alesci:2013xd}. Here we review the procedure, pointing out all the assumptions, which we can classify into two categories: i) an additional modification (put by hands) not being a standard method of a field theory or LQG; ii) an internal gauge fixing introduced as an additional constraint at the quantum level. We also emphasize which steps in this method we consider to be incorrect. Finally, we distinct a reduction of states that we label by sub-point a) and a reduction of operators labeled by b).

\begin{enumerate}[I)]
\item\label{rI}
The first assumption in this method is solving the Gauss constraint in \eqref{A_Gauss}, quantized and imposed on $\mathcal{H}_{kin}^{(gr)}$. As a result, we obtain the Gauss-invariant Hilbert space of LQG, already discussed in Sec.~\ref{II.4}. This is a standard procedure in the theory, but we placed it here, since it is a modification of the unconstrained Hilbert space before the next steps of reduction take place --- these have to be done before solving HCO. It is worth mentioning that the order of resolution of the vector constraint operator would be also influential into this analysis. However, it is not explicitly written in  \cite{Alesci:2013xd} whether this operator is solved after or before the reduction steps listed below. Anyhow, all three first class constraints as the elements of the standard Dirac's method of quantization \cite{Dirac:1950pj}, should be solved one by one without any intermediate modification. Alternatively, one can use one of the gauge-fixing methods listed in Appendix \ref{D}, introducing canonical gauge conditions, constituting a second class constraint system with one or few of the original first class constraints. Therefore we already found this first step to be extremely problematic. One cannot first solve the constraint and later impose gauge conditions, since these do not Poisson-commute with the original constraints, as shown in appendix E. The imposed constraint is (by definition of the second class system) not invariant under the symmetry introduced by the gauge conditions. This is exactly the problem that appears in this reduction procedure, leading to an ill-defined Hilbert space.
\item\label{rII}
The next additional modification of LQG is a restriction on $\Gamma$ to be cuboidal,
\begin{align}\label{cuboidal}
\Gamma\ \to\ {}^R\Gamma\,.
\end{align}
This is a cosmologically motivated simplification, which does not commute with the diffeomorphism constraint. It is however not clear whether it reduces any gauge transformation completely or not --- three rigid diffeomorphism transformations are still present. Thus, some restriction on the Hilbert space is introduced, however not in a controllable way. Therefore after this step, one will not be able to detect whether the final result is correct or not. It is also worth mentioning that this restriction has no influence on the canonical operators, but it restricts loop holonomies and densitized dreibeins to act along rigid directions. Hence, only the variables in \eqref{QRLG_A} and \eqref{QRLG_E} contribute in the definition of the operators. Once the cuboidal symmetry has been assumed, this must be preserved. This prevents from acting with holonomy operators along directions that do not coincide with the edges of the cuboidal graph. Such an action, if considered, would indeed lead us out of the Hilbert space of the reduced theory.   
\item\label{rIII}
The third externally introduced assumption is a freezing of the internal symmetry. The $\mathfrak{su}(2)$ generators become fixed along the directions of the ${}^R\Gamma$ lattice, namely $\tau^p=\text{P}^p_i\tau^i$. Notice that the internal rotation operator $\text{P}^p_i$ has to be fixed with respect to the orthogonal Cartesian frame spanned by $t^i$. Therefore, the translational invariance of the Ashtekar variables in \eqref{QRLG_A} and \eqref{QRLG_E} with the fixed spatial directions, $x^p$, has to be fixed as well. A change in the scaling of the spatial directions would affect the structure of the $\text{P}^p_i$ operator (and the choice of the $\vec{\rho}$ vector --- compare with Appendix \ref{B}).
\begin{enumerate}[a)]
\item\label{rIIIa}
The basis states of the reduced spin-network are then chosen to be the U$(1)_p$-invariant, diagonalizing holonomies in \ref{rIIIb}).
\item\label{rIIIb}
The reduced holonomies take the following general form, $h_p=e^{\alpha_{(p)}\tau^p}$. Acting on states in \ref{rIIIa}), they result in eigenvalue $e^{-\mathsf{i}\alpha_{(p)}m^{(p)}}$, where $\alpha_{(p)}$ is a function of the Ashtekar connection and of a link, while $m^{(p)}$ is the magnetic quantum number corresponding to the spin attached to the $l^p$ link --- see Appendix \ref{A}.
\end{enumerate}
This step reduces the symmetry of the solutions at the quantum level. However any degree of freedom is removed, since the SU$(2)$ symmetry has been replaced with the $\bigotimes_p^3\text{U}(1)_p$ one. This step, if subsequent to \ref{rI}), would be the only correct one. But, if the order is reversed as here, restriction on the graph in \eqref{cuboidal} cannot be consistently attained. To clarify this argument, one cannot rotate gauge generators without changing intertwiners that were already affixed with respect to the generators along the original directions.
\item\label{rIV}
Another modification of the theory is an introduction of the SU$(2)$ projected spin-network by lifting up the reduced U$(1)_p$ state space (for a general idea of projected spin-networks see \cite{Dupuis:2010jn}).
\begin{enumerate}[a)]
\item\label{rIVa}
The projected states are constructed by an extension of the U$(1)_p$ ones to the subsector of SU$(2)$ restricted to generators $\tau^p$. This can be done by convolutions of $\bigotimes_p^3\text{U}(1)_p$ and SU$(2)$ characters.
\item\label{rIVb}
Notice that, according to the construction of the projected spin-networks in \cite{Dupuis:2010jn}, `gluing' subspaces in order to lift them into the ones associated to a more general symmetry, should not affect the matrix elements of the operators. Therefore, at this step, we keep unchanged the reduced form of holonomies given in Appendix \ref{B}.
\end{enumerate}
Here, again the problem of the order of steps occurs. One should first choose the cuboidal lattice, then impose the Gauss constraint (on the already diagonal directions), finally lifting up $\bigotimes_p^3\text{U}(1)_p$ to the subsector of SU$(2)$. This `trick' however would not then change anything, since generators of both symmetries span the same space.
\item\label{rV}
\begin{enumerate}[a)]
\item\label{rVa}
Finally, considering now the lifted SU$(2)$ state space, one can solve the following gauge condition,
\begin{align}
\label{internal_gauge}
\chi_i\chi_i=E^a_j\,{}^{0\!}e_a^k\big(E^b_j\,{}^{0\!}e_b^k-E^b_k\,{}^{0\!}e_b^j\big)=0\,.
\end{align}
Solving it at the quantum level, \textit{i.e.} replacing densitized dreibeins with flux operators
\begin{align}
\label{dd_operator}
\hat{E}^i(\mathbf{S}):=\!\int_{\!S}\!dx^adx^b\epsilon_{abc}\hat{E}_i^c\,,
\end{align}
and identifying the internal directions with the lattice ones, we ensure that the SU$(2)$-fixing constraint,
\begin{align}
\label{second_class}
\chi_i:=\epsilon_{ijk}\,{}^{0\!}e_a^jE^a_k=0\,,
\end{align}
is implemented weakly. Let us also specify that \eqref{second_class} constitutes a second-class constraint system with the Gauss constraint in \eqref{A_Gauss} --- see \cite{Cianfrani:2011wg,Cianfrani:2012gv} for details. Therefore, to omit the procedure involving Dirac brackets, expression \eqref{internal_gauge} has been introduced as a `master constraint', containing two elements: $E^a_jE^b_j\,{}^{0\!}q_{ab}$ and $E^a_jE^b_k\,{}^{0\!}e_a^k\,{}^{0\!}e_b^j$, of which the former one Poisson commutes with the Gauss constraint and the latter one generates with it the same algebra as \eqref{second_class}. Notice that in this way we break the SU$(2)$ group of rotations into U$(1)$ in a controlled way, introducing another constraint on the full theory.
\\
\hspace*{4mm}Solutions to the equation $\chi_i\chi_i=0$ derived at the quantum level (see \cite{Alesci:2013xd} for the details) are realized by the states corresponding to the quantum numbers satisfying the relations
\begin{align}\label{spin_solution}
j^p\ \to\ \infty
\intertext{and}
\label{magnetic_solution}
m^p\simeq\bar{m}^p\,.
\end{align}
Therefore, when defining the reduced spin-network, the limit in \eqref{spin_solution} is approximated by \eqref{spin_number}, while the approximation in \eqref{magnetic_solution} becomes simplified by the sharp equality, as in \eqref{magnetic_number}. This means that all the states along three orthogonal directions are no longer different, which introduces homogeneity, preserving the anisotropy of the model.
\\
\hspace*{4mm}It is worth noting that although this reduction step is implemented only with respect to the states, it was introduced as a new constraint, thus it should impose a restriction on the states space, with no influence on the operators. This interpretation could prevent from the argument of breaking a consistency between states and operators, while implementing the modification. What is problematic however is the question whether the homogeneity of states should entail homogeneity of operators.
\item\label{rVb}
In QRLG operators remain in the inhomogeneous form. We consider this choice as inconsistent in the implementation of the symmetry within the theory. This is because inhomogeneous operator acting on a homogeneous state creates a new, inhomogeneous state, which should still preserve the symmetry introduced by the gauge condition.
\end{enumerate}
The problem that arises here is very serious. Condition \eqref{internal_gauge} develops the second class set with all the constraints, \eqref{A_Gauss}, \eqref{A_vector} and \eqref{A_scalar} --- see Appendix \ref{E}. Thus it is no easy to detect which degrees it annihilates. Moreover, it coincides with the weakly implemented condition \ref{second_class}. The latter one definitely annihilates three degrees of freedom, thus no dynamics will be left out of the phase space spanned by the variables in \eqref{QRLG_A} and \eqref{QRLG_E}, while the scalar and (probably) the vector constraints have not been implemented yet. Due to the many flaws occurring during the previous steps, from now on the theory has no propagating degrees of freedom, while there are still some degrees to be subtracted.
\item\label{rVI}
The vector constraint is probably implemented at this moment. We deduced this, because the symmetry restricted by the vector constraint has been derived in \cite{Alesci:2013xd} after the Gauss constraint has been completely implemented and reduced. Diffeomorphisms have been then restricted to the translations along rigid Cartesian directions, $x^a\to x^p=x^a\,^{0\!}e^p_a$. This corresponds to the variables in \eqref{R-C_A} and \eqref{R-C_E}, instead of \eqref{QRLG_A} and \eqref{QRLG_E}, respectively.
\item
\begin{enumerate}[a)]\setcounter{enumii}{1}
\item\label{rVIIb}
The last assumption is to replace the rotated holonomies obtained in \ref{rIIIb}) with general ones, restricted only by the rotation of generators into $\tau^p$. In this way the mechanism for lifting up the U$(1)_p$ states into SU$(2)$ ones in \ref{rIVa}) is extended into the operators. Notice that without any changes, this step can be also implemented just after point \ref{rIV}). However, this additional modification is not a consequence of the projected spin-network technique, and thus it results in a much wider spectrum of operators than in the case we discussed in \ref{rIVa}). In other words, the problem in this step is rather methodological than mathematical: reduction on the states space would not entail reduction of the operators (namely restriction to the subset of operators the image of which is inside the states space). 
\end{enumerate}
\end{enumerate}

Summarizing, the procedure introduced in \cite{Alesci:2013xd} appears to be complicated, mixing gauge fixing methods from Appendix \ref{D} with themselves and with non-standard techniques. Moreover, as we pointed out, few incorrect steps were considered. However, in \cite{Alesci:2013xya} a simpler method, which we discuss in the next subsection, was proposed.

\subsection{Reduction procedure II: double gauge conditions in QRLG}\label{IV.2}

\noindent
The reduction scheme described in the previous subsection can be simplified and performed in a more controlled way. In the original construction, four additional modifications and one gauge condition are needed. Here, referring to the method described in \cite{Alesci:2013xya}, we obtain the same results using two constraints and two additional modifications.

\begin{enumerate}[1)]
\item\label{r1}
We begin from the same problematic assumption as in \ref{rI}), first solving the Gauss constraint.
\item
\begin{enumerate}[a)]
\item\label{r2a}
Next, we introduce a new gauge condition that classically describes the vanishing of the off-diagonal entries of the dreibein (and densitized dreibein) matrices,
\begin{align}
\label{external_gauge}
E^a_iE^b_i\big|_{a\neq b}\!=0\,.
\end{align}
Notice that this constraint is introduced as a global condition, breaking the SO$(3)$ invariance of the $q_{ab}$ spatial metric, and leads to Ashtekar variables that in the most general form can be represented by expressions \eqref{QRLG_A} and \eqref{QRLG_E}. This gauge condition Poisson-commutes with the Gauss constraint (see \eqref{Gauss_eta}) and creates a second class set only with the vector and scalar constraints --- see Appendix \ref{E}. Thus implementing the first step \ref{r1}) seems so far to be correct.
\\
\hspace*{4mm}At the quantum level, this condition leads to the large-$j$ limit solution in \eqref{spin_solution} and the cuboidal graph structure \eqref{cuboidal}. Here, likewise in subsection \ref{IV.1}, it is not specified whether the implementation at the quantum level of the new constraint in \eqref{external_gauge} affects only the states, or it also entails a similar reduction of the operators. Moreover, analogously to the procedure in \ref{IV.1}, this step becomes incorrect at the quantum level, since the Gauss constraint has been already solved in step \ref{r1}). The intertwiners derived for non-orthogonal links are in contradiction to the restriction in \eqref{cuboidal}.
\end{enumerate}
\item\label{r3}
Then, we fix the holonomies and the states, respectively, to the ones with frozen generators $\tau^p$, and to the ones carrying eigenvalues of the holonomies. This is a similar step to the one discussed in \ref{rIII}). It is worth mentioning that this procedure was introduced by hand and unnecessarily, since in stead this should already be a consequence of a correct implementation of the Gauss constraint in \ref{r1}). 
\item\label{r4}
The fourth condition is again a gauge fixing one, the same as in \ref{rV}). The only difference with respect to the method in subsection \ref{IV.1}, concerns the implementation of the gauge fixing to U$(1)$, not on the SU$(2)$-invariant space, but on the $\bigotimes_p^3\text{U}(1)_p$ states space.
\item
When the condition introduced above is solved, the gauge condition in \eqref{external_gauge} restricts the vector constraint to be a generator only of the reduced diffeomorphisms --- we discussed this in the previous procedure, in \ref{rVI}). Again, we put this step not together with \ref{r2a}), but after \ref{r4}), since also the gauge condition in \eqref{internal_gauge} constitutes a second class system with the vector constraint. If the Gauss constraint would be implemented correctly, we would be able to implement all the first class constraints in any order.
\item
\begin{enumerate}[a)]\setcounter{enumii}{1}
\item\label{r6b}
Finally, in this simplified construction of QRLG, the additional modification, already described in \ref{rVIIb}), is introduced. Hence again holonomies are promoted into general SU$(2)$ operators, while states are not changed in an analogous way.
\end{enumerate}
\end{enumerate}

Concluding the enhanced reduction procedure, the number of external modifications to the theory, as well as incorrect implementations of constrains and gauge conditions has been lowered. However, few problematic steps as in Sec.~\ref{IV.1} still appear in Sec.~\ref{IV.2}. Especially, the theory remains overconstrained --- see discussion in Sec.~{IV.3}.

\subsection{Discussion}\label{IV.3}

\noindent
We already pointed out the problematic steps in the gauge fixing procedure of QRLG. Let us analyze now directly the two troublesome consequences of the incorrect gauge reduction.

First, let us discuss the presence of the intertwiners in the final result. As the authors wrote in the review of their model, ``the presence of intertwiners is the main technical achievement of QRLG'' \cite{Alesci:2016gub}. In our opinion, the objects removing the orthogonality of states, hence resulting in an ill-defined Hilbert space, should be rather the main reason to reinvestigate the procedure, in the attempt of finding the solution of the problem. It is easy to see that, in contradiction of the Henneaux and Teitelboim recipe for a precise reduce phase space quantization --- sketched in Appendix \hyperref[D.3]{D.3} --- the reduced state, containing intertwiners linking non-orthogonally positioned gauge generators, does not coincide with the orthogonality restriction imposed on the gauge invariant spin-network state (as we stated in the previous sections). The proper state would be rather linking the same generators along a (straight) line --- thus with trivial, negligible intertwiners.

Second, QRLG is presented as a model that after an unclear gauge fixing procedure results in a six-dimensional phase space spanned by the variables in \eqref{QRLG_A}, \eqref{QRLG_E}, while all the original constraints of the full theory remains present (although in a simplified form). Let us recall the reduced form of the Gauss constraint,
\begin{align}
\label{Ared_Gauss}
{}^RG^{(A)}=&\;\frac{1}{\gamma\kappa}\!\int_{\Sigma_t}\!\!\!\!d^3x\,A^i_t\,\partial_a{}^R\!E^a_i\,,
\intertext{and the reduced diffeomorphism constraint,}
\label{Ared_vector}
{}^RV^{(A)}
=&\;\frac{1}{\gamma\kappa}\!\int_{\Sigma_t}\!\!\!\!d^3x\,N^a\big(\partial_a{}^R\!A_b^i-\partial_b{}^R\!A_a^i\big){}^R\!E^b_i\,.
\end{align}
The former expression corresponds to $\propto\partial_pp^{(p)}(t,x^{(p)})$, and generates the U$(1)_p$-invariance. The constraint in \eqref{Ared_vector} simplifies to $\propto\partial_pc^{(p)}(t,x^{(p)})$ --- the terms $\propto\partial_pc^{(q)}|_{p\neq q}$ vanish. Imposing the Gauss constraint, $\partial_pp^{(p)}=0$, and neglecting the boundary term, the reduced diffeomorphism constraint can be recast as an expression proportional to $\propto\partial_pN^{(p)}$ --- see also an alternative explanation in \cite{Alesci:2013xd}. This leads to the conclusion that the shift vector, as a generator of the reduced diffeomorphism, reveals a dependence only on the fixed lattice directions $N^{(p)}(t,{\bf x})=N^{(p)}(t,x^{(p)})$. Hence, diffeomorphism transformations are restricted to the translations along the three directions of the links of ${}^R\Gamma$.

These constraints are present in the earlier formulation of the theory in \cite{Alesci:2013xd}. Adding then the scalar constraint, results in $(3+3+1)\cdot2=14$ dimensions to be subtracted from the six-dimensional phase space.

As we discussed in Sec.~\ref{IV.2}, it is not clear how the Gauss constraint is implemented in the enhanced formulation of QRLG in \cite{Alesci:2013xya}. From the analysis above, one may guess that it still remains as a constraint equation, however we are not going to defend this statement. Hence, let us assume that only the reduced diffeomorphism constraint remains present. This anyhow keeps the theory overconstrained, since eight dimensions must be subtracted out of six-dimensional phase space.


\section{Standard reduction of kinematics of LQG}\label{V}

\noindent
We proved that the Alesci-Cianfrani model is incorrect. Let us propose now the simplest correction to the reduction procedure, using the gauge conditions introduced in QRLG.

\subsection{Reduction procedure III: reduced phase space}\label{V.1}

\noindent
As it has been demonstrated in Appendix \ref{E}, the six gauge conditions proposed by Alesci, Cianfrani and Rovelli constitute a second class system, each with higher number of constraints. Thus they do not fix any symmetry completely. To check the result of their implementation, let us follow the procedure sketched in Appendix \hyperref[D.3]{D.3}.

The global solution to the three conditions
\begin{align}
\label{chi_E}
\chi_i=\epsilon_{ijk}{}^{0\!}E^j_k=0\,,
\end{align}
is not as claimed by the inventors of QRLG --- a diagonal form of the dreibein density, but is rather provided by the relation 
\begin{align}
\label{chi_S}
{}^{0\!}E^i_j={}^{0\!}E^j_i\,.
\end{align}
Here we keep the notation introduced in Appendix \ref{E}, defining ${}^{0\!}E^i_j:={}^{0\!}e^i_aE^a_j$. This correctly reduces exactly the same number of degrees of freedom as the number of gauge conditions. We can then specify the form of $\chi$-reduced symmetric dreibein density as follows,
\begin{align}
\label{chi_E_solved}
{}^{0\!}E^i_j(t,{\bf x})\to{}^{\,0\!}_{\chi\!}E^i_j(t)=\left(
\begin{array}{ccc}	{}^{0\!}E^1_1	&	{}^{0\!}E^1_2	&	{}^{0\!}E^1_3	\\
				{}^{0\!}E^1_2	&	{}^{0\!}E^2_2	&	{}^{0\!}E^2_3	\\
				{}^{0\!}E^1_3	&	{}^{0\!}E^2_3	&	{}^{0\!}E^3_3	\end{array} \right).
\end{align}
The homogeneity of ${}^{\,0\!}_{\chi\!}E^i_j(t)$ guarantees that its form remains preserved globally. Thus that the gauge condition in \eqref{chi_E} is free from the Gribov obstruction --- see Appendix \hyperref[D.3]{D.3}.

Solving next the three conditions 
\begin{align}
\label{eta_E}
\eta^{ab}=E^a_iE^b_i\big|_{a\neq b}=0\,,
\end{align}
one needs to derive the set of three equations,
\begin{align}
\label{eta_S}
\begin{split}
{}^{0\!}E^1_1{}^{0\!}E^1_2+{}^{0\!}E^1_2{}^{0\!}E^2_2+{}^{0\!}E^1_3{}^{0\!}E^2_3=&\;0\,,
\\
{}^{0\!}E^1_1{}^{0\!}E^1_3+{}^{0\!}E^1_2{}^{0\!}E^2_3+{}^{0\!}E^1_3{}^{0\!}E^3_3=&\;0\,,
\\
{}^{0\!}E^1_2{}^{0\!}E^1_3+{}^{0\!}E^2_2{}^{0\!}E^2_3+{}^{0\!}E^2_3{}^{0\!}E^3_3=&\;0\,.
\end{split}
\end{align}
We are looking for the solution, keeping any three components of ${}^{\,0\!}_{\chi\!}E^i_j$ unconstrained. The unique solution fulfilling this requirement reads,
\begin{align}
\label{constrained_E}
{}^{0\!}E^1_2={}^{0\!}E^1_3={}^{0\!}E^2_3=0\,,
\end{align}
with diagonal elements of ${}^{0\!}E^i_j$, being the only non-vanishing degrees of freedom\footnote{Notice that conversely to the methodology of QRLG, we do not claim that diagonalization of the densitized dreibein entails diagonalization of the Ashtekar connection. Moreover, what we specified in Appendix \ref{D}, fixation of any gauge condition has to be done globally, hence the results of QRLG in \eqref{QRLG_E} or \eqref{R-C_E} could not be the solution of the gauge conditions.}.

The reduced densitized dreibein can be then written in the form analogous to the one introduced by LQC \cite{Ashtekar:2003hd,Ashtekar:2009vc},
\begin{align}
\label{QRLGh_E}
\bar{E}_{i}^a(t):=\frac{l_0^{(i)}}{{\bf V}_{\!0}}\bar{p}^{(i)\!}(t)\,\sqrt{^{0\!}q}\,{}^{0\!}e_{i}^a\,.
\end{align}
Notice that due to the fact that the gauge conditions Poisson-commute, see \eqref{chi_eta}, reversing the order of their implementation, the final result remains unchanged.

Implementing then the diagonal dreibein solution to the constraint equations in \eqref{A_Gauss}, \eqref{A_vector} and \eqref{A_scalar}, we realize that none of them is vanishing. This verifies that the gauge conditions have been chosen correctly: symmetries have been globally fixed, while the number of constraint equations remains unchanged. In other words, we traded the six degrees of freedom, encoded in the orbits of SU$(2)$ and in the spatial diffeomorphisms symmetries, to reduce the dimensions of the phase space by six.

There is, however, a problem that we must solve in order to restore the correct rules of canonical quantization. So far, the standard correspondence principle, leading to the commutation relation
\begin{align}
\Big[\hat{A}^i_a(t,{\bf x}),\hat{\bar{E}}_{i}^a(t)\Big]=\i\kbar\delta_a^b\,\delta_j^i\,\delta^{(3)\!}\big({\bf x}\big)
\quad
\text{(inconsistent),}
\end{align}
where $\kbar:=\gamma\hbar\kappa/2$, would be in contradiction to the Poisson bracket in \eqref{Poisson_LQG} for any non-diagonal pair of the canonical variables. Locality of the operator $\hat{A}^i_a$ creates another difficulty.

The easiest way to overcome this problem is to follow the reversed method to the one sketched in Appendix \hyperref[D.2]{D.2}. We can simply redefine the correspondence rules of quantization as follows,
\begin{align}
A^i_a(t,{\bf x})\to
\left\{\begin{array}{ll}
\hat{\bar{A}}^{i}_a(t):=\frac{1}{l_0^{(i)}\!}\hat{\bar{c}}_{(i)\!}(t){\,}^{0\!}e^{i}_a
\quad
&\text{ for }{}^{0\!\!}A^j_i\propto\delta^j_i\,,
\\
0
&\text{ for }{}^{0\!\!}A^j_i\big|_{i\neq j}\!=0\,,
\end{array}\right.
\end{align}
where analogously to \eqref{projected_E}, we introduced the Ashtekar connection projected along fiducial directions,
\begin{align}
\label{projected_A}
{}^{0\!\!}A^j_i:={}^{0\!}e^a_iA^j_a\,.
\end{align}
Then, since the image of the correspondence rules is isomorphic to the quotient of the space, spanned by the variable $A^i_a$, by the kernel of these rules, we can simply restrict the classical space to this quotient. To keep the notation specified in \eqref{QRLGh_E}, we define
\begin{align}
\label{QRLGh_A}
\bar{A}^{i}_a(t):=\frac{1}{l_0^{(i)}\!}\bar{c}_{(i)\!}(t){\,}^{0\!}e^{i}_a\,.
\end{align}
To ensure that we reduced the phase space correctly, we verify that the restriction to the homogeneous, anisotropic variables in \eqref{QRLGh_E} and \eqref{QRLGh_A}, entails the vanishing of each one of the three Gauss equations and of the three vector constraints. Let us remind that this happened after removal of the six components of the Ashtekar connection, ${}^{0\!\!}A^j_i\big|_{i\neq j}\!=0$, preserving the initial number of propagating degrees of freedom of LQG.

The method above appears to be correct. To be certain, however, we should rather directly solve the Gauss and vector constraints expressed in terms of the reduced densitized dreibein in \eqref{QRLGh_E}. The former constraint equation seems to be simpler, leading to the following global condition,
\begin{align}
\label{Gauss_A}
\epsilon_{ijk}A^j_a\bar{E}_{i}^a=0\,.
\end{align}
This is an analogous equation to the gauge condition in \eqref{chi_E}, thus providing a solution analogous to \eqref{chi_E_solved}.
Solving globally the vector constraint equation, we use the diagonality of ${}^R\!\bar{E}_{i}^a$ and the spatial homogeneity of ${}^{0\!\!}A^i_j={}^{0\!\!}A^j_i$, which lead to the set of three relations,
\begin{align}
\label{vector_A}
\epsilon_{jkl}{}^{0\!\!}A^j_i{}^{0\!\!}A^k_l=0\,.
\end{align}
These equations are identically fulfilled for diagonal connections, while for the non-diagonal ones we get
\begin{align}
\label{constrained_A}
{}^{0\!\!}A^j_i\big|_{i\neq j}\!=0\,.
\end{align}
This verifies the result in \eqref{QRLGh_A}.

\subsection{Reduction procedure IV: quantum gauge conditions}\label{V.2}

\noindent
Let us repeat now the reduced phase space procedure at the quantum level. Conversely to the incorrect methods of QRLG, we solve both operator equations in the same manner. In this procedure, we solve them directly.
\begin{enumerate}[a)]
\item
\label{ra}
First, we introduce the gauge fixing equation, analogous to the condition \eqref{external_gauge} or \eqref{eta_E},
\begin{align}
\label{quantum_eta_E}
\hat{E}_{i\!}\big(\mathbf{S}^{(p)}\big)\hat{E}_{i\!}\big(\mathbf{S}^{(q)}\big)\big|_{p\neq q}|\Gamma;J,I\rangle=0
\ \Leftrightarrow\ 
\forall_{v,j^l}\hat{E}_{i\!}\big(\mathbf{S}^{(p)}\big)\hat{E}_{i\!}\big(\mathbf{S}^{(q)}\big)\big|_{p\neq q}|v;j^l\rangle=0
\,.
\end{align}
As presented in \cite{Alesci:2013xya}, global solution of this operator equation selects the states of the cuboidal graphs --- we discussed this in Sec.~\ref{IV.2}. An important remark here is that the related directions of the eigenvalues of the operators $\hat{E}_{i\!}\big(\mathbf{S}^{(p)}\big)$ are distinguishable in the large-$j$ spin numbers limit \cite{Bianchi:2009ri}.
\item
\label{rb}
Let us consider the cuboidal graphs, being a solution of the previous step. We restrict then the second gauge fixing equation only to the orthogonal directions (which is however not necessary to obtain the results below). The action of the flux operator leads to the following eigenvalue,
\begin{align}
\hat{E}_{k\!}\big(\mathbf{S}^{(q)}\big)h_{l^p}
=\pm\i\kbar\,h_{l^p_{\text{in}}}\tau^kh_{l^p_{\text{fin}}}\delta^{(q)}_{(k)}\,,
\end{align}
where $h_{l^p_{\text{in}}}\!h_{l^p_{\text{fin}}}\!=h_{l^p}$. Then the quantized condition \eqref{second_class} or \eqref{chi_E} acting along any link $l\in{}^R\Gamma$ reads,
\begin{align}
\label{quantum_chi_E}
\epsilon_{ijk}{}^{0\!}e_{(q)}^j\hat{E}_{k\!}\big(\mathbf{S}^{(q)}\big)h_{l^p}=0\,.
\end{align}
This equation is always satisfied for $j=q=k$.
\end{enumerate}
Notice that the order of implementation of the conditions above is exchangeable: independently of links directions, condition \eqref{quantum_chi_E} is identically satisfied for $j=q=k$, while \eqref{quantum_eta_E} can be solved for operators $\hat{E}_{i\!}\big(\mathbf{S}^{(i)}\big)$.

\subsection{Reduction procedure V: squared quantum conditions}\label{V.3}

\noindent
Let us follow now the method proposed in QRLG and define --- as it is suggested there --- the master constraints, as squares of the operators in \eqref{quantum_eta_E} and \eqref{quantum_chi_E}. Conversely to the procedure in the Alesci-Cianfrani theory, for consistency we consider to do it for both quantum conditions.

\begin{enumerate}[A)]
\item
\label{rA}
The quantum realization of the condition $\eta^{ac}q_{cd}\eta^{db}=q\eta^{ab}$ reads
\begin{align}
\label{master_eta_E}
\hat{\mathbf{V}}^{2}\hat{E}_{i\!}\big(\mathbf{S}^{(p)}\big)\hat{E}_{i\!}\big(\mathbf{S}^{(q)}\big)\big|_{p\neq q}|\Gamma;J,I\rangle=0\,.
\end{align}
Independently on the structure of the node on which the operators act, this equation must be satisfied globally. Therefore, this condition reduces to the one in paragraph \ref{ra}) in the previous subsection, and it is unrelated to the eigenvalue of the volume operator.
\item
\label{rB}
The square master equivalent of \eqref{quantum_chi_E} has been already discussed in the previous section, \ref{rVa}). The operator equation,
\begin{align}
\label{master_chi_E}
{}^{0\!}e_{(q)}^j\hat{E}_{k\!}\big(\mathbf{S}^{(q)}\big)\Big[
{}^{0\!}e_{(r)}^j\hat{E}_{k\!}\big(\mathbf{S}^{(r)}\big)-{}^{0\!}e_{(r)}^k\hat{E}_{j\!}\big(\mathbf{S}^{(r)}\big)
\Big]h_{l^p}
=-\kbar^2\,{}^{0\!}e_{(k)}^j\!\Big[
\tau^k\tau^kh_{l^p}{}^{0\!}e_{(k)}^j
-h_{l^p_{\text{in}}}\tau^k\tau^jh_{l^p_{\text{fin}}}{}^{0\!}e_{(j)}^k
\Big]=0\,,
\end{align}
vanishes for any link direction if $j=q=k$, in the same manner as in paragraph \ref{rb}) in the previous subsection.
\end{enumerate}

We demonstrated that the squared quantum conditions did not change the structure of the reduced Hilbert space. Since the reduction must be implemented globally on every state, it must be implemented also on the states modified by an action of a holonomy operator. Therefore, the holonomy operators have to be restricted to orthogonal directions of ${}^R\Gamma$, with $\mathfrak{su}(2)$ generators affixed along these directions.

One may ask whether the space of the reduced states is a subspace of the kinematical Hilbert space of LQG. The answer is positive, and the map to the reduced subspace has been already explained in \ref{rIII}) and \ref{r3}) --- generators are rotated, simplifying the structure of the Wigner $D$-matrices (see Appendix \ref{B}).

Finally notice that reducing the theory both at the classical and at the quantum level leads to the same structure of the Hilbert space and to the same canonical operators. It is however unclear whether the remaining constraint operator --- the Hamiltonian constraint --- has identical spectrum. The action of the Lorentzian term, proportional to the expression in \eqref{curvature}, has not been derived yet in the anisotropic cosmological framework\footnote{Notice that this problem has been totally neglected in QRLG. Ignoring these flaws of the model leads to a significantly different quantum theory than the anisotropic extension of LQC. The derivation of the Lorentzian operator would be then necessary in order to discuss any phenomenological aspects of the theory.}. We leave this issue as an open problem, with the solution being under construction \cite{Bilski_Lorentzian}.


\section{Sketch of kinematics of reduced LQG}\label{VI}

\noindent
Introducing the corrected formulation of the reduction procedure, LQG has become a simple symmetry-frozen theory. Let us present here only the sketch of the quantum reduction of canonical gravity, while all the details will be explained in a shortly coming article \cite{Bilski_CQC}.

The states realizing the results of the quantum reduction LQG presented in \ref{V.2} and \ref{V.3} are defined as follows,
\begin{align}
\label{simplified_state}
\langle\bar{h}|\Gamma;J\rangle
:=\prod_{l^p\in\Gamma}\!^{p}\!D^{|\bar{m}|^p}_{\!\bar{m}^p\,\bar{m}^p}(\bar{h}_p)\,,
\end{align}
where $^{p}\!D^{|\bar{m}|^p}_{\!\bar{m}^p\,\bar{m}^p}(\bar{h}_p)$ are the same Wigner $D$-matrices as in the definition in \eqref{unnormalized}. These states span a well-defined Hilbert space, with an orthogonality condition given by the scalar product
\begin{align}\label{simplified_scalar_product}
\big<\Gamma;\bar{m}^{l^p}\big|\Gamma'\!,\bar{m}^{{l'}^q}\big>
=\delta_{\Gamma,\Gamma'}\prod_{l\in\Gamma}\delta_{j^{l^p}\!\!,\,j^{{l'}^q}}\,.
\end{align}

Rewriting the total Hamiltonian in \eqref{A_total_Lambda} in terms of the homogeneous variables defined in \eqref{QRLGh_A} and \eqref{QRLGh_E}, considerably simplifies the dynamics. Indeed, this takes the form
\begin{align}
H^{(\bar{A})}_T+H^{(\Lambda)}_T
=H^{(A)}+H^{(\Lambda)},
\end{align}
while the constraints in \eqref{A_Gauss} and \eqref{A_vector} vanish identically.

All the results of the original QRLG concerning flux-dependent operators remain unchanged. For instance the eigenequation of the reduced flux operator acting at a uniquely defined point $l^{(q)}\cap\mathbf{S}$, reads
\begin{align}
\begin{split}
\label{simplified_flux}
\hat{\bar{E}}^p(\mathbf{S})
\big|l^{(q)}\!,\bar{m}^{(q)}\big>
=&\;-\i\kbar\!\int_{\!S}\!dx^sdx^t\epsilon_{rst}\frac{\delta}{\delta\bar{A}^p_r(x^{(r)})}
\big|l^{(q)}\big(y^{(q)}\big),\bar{m}^{(q)}\big>
\\
=&\;-\i\kbar\delta^{(q)}_p
\big<\tau^3\big|l^{(q)}\!,\bar{m}^{(q)}\big>
=-\kbar\bar{m}^{(q)}\delta^{(q)}_p
\big|l^{(q)}\!,\bar{m}^{(q)}\big>\,.
\end{split}
\end{align}
It is worth mentioning that the difference in sign with respect to the Alesci-Cianfrani formulation in \cite{Alesci:2013xd,Alesci:2013xya,Alesci:2014uha,Alesci:2016gub,Alesci:2017kzc} comes from the initial choice of the $\tau^i$ generators and the real holonomies --- we already discussed that in Sec.~\ref{II.3}.

Notice also that since a holonomy operator acts on states as a multiplication, the aforementioned eigenequation is the only expression that imposes significant constraints on the structure of the states. Moreover, it is easy to see that formula \eqref{simplified_flux} has the form of a simple differential equation, with the solution
\begin{align}
\label{simplified_basis}
\big|l^{(q)}\!,\bar{m}^{(q)}\big>=\exp\!\bigg(\!-\i \bar{m}^{(q)}\!\!\int\!\!dx^u\bar{A}^{(q)}_u\bigg).
\end{align}
Thus, without any loss of generality, we may consider this expression as a definition of a state. As demonstrated in the Appendix \ref{B}, the space of states constructed from the basis states above is a subspace of the kinematical Hilbert space of LQG\footnote{ Precisely, the Fock space is defined as the sum over all the nodes of the states $\bigotimes_q\sum_{\bar{m}^{(q)}}\alpha_{\bar{m}^{(q)}}\big|l^{(q)}\!,\bar{m}^{(q)}\big>$, where $\forall_{\bar{m}^{(q)}}\alpha_{\bar{m}^{(q)}}\in\mathds{C}$. It is also worth mentioning that it has been demonstrated that this states space can be embedded in the Fock space of LQG \cite{Engle:2007zz,Brunnemann:2007du,Engle:2013qq,Bodendorfer:2015hwl}.}.

Let us now derive the action of a square of the volume operator defined on an open neighborhood $B_{l_0\!}(v)$,
\begin{align}\label{simplified_volume}
\big(\hat{\mathbf{V}}_{\!v,l_0}\big)^{\!2}
|\Gamma;J\rangle
=
\kbar^3\prod_p^3\!\Big|\bar{m}^{(p)}(v)+\bar{m}^{(p)}\big(v-l_0^{(p)}\big)\Big|
|\Gamma;J\rangle\,.
\end{align}
Here, $v\in\Gamma$ is a hexavalent node, \textit{i.e.} there are six links emanated from $v$, while $v-l_0^{(p)}$ denotes a nearest neighbor node in a negatively oriented $p$ direction at a distance $l_0^{(p)}$. Operators $\hat{E}_i({\bf S}^p)|_{i=p}$ from expression \eqref{volume_operator} act at $l^{(p)}(v-l_0^{(p)})\cap\mathbf{S}_-$ and at $l^{(p)}(v)\cap\mathbf{S}_+$ --- the vertices in the brackets represent initial points of the positively oriented collinear links connected at $v$. Notice also that to derive the eigenvalue of $\hat{\mathbf{V}}_{\!v,l_0}$, we used the eigenvalue of the flux operator in \eqref{simplified_flux}, assuming an unidirectional orientation of a collinear link frame $y^{(q)}$ and a coordinate frame $x^{(r)}$.

It is worth mentioning that the derivation of the square root of the result in \eqref{simplified_volume} would be problematic. However, the root of an analogous quartic matrix element is well defined,
\begin{align}\label{simplified_matrix_volume}
\sqrt[4]{\langle|\Gamma;J|\big(\hat{\mathbf{V}}_{\!v,l_0}\big)^{\!4}|\Gamma;J\rangle}\,.
\end{align}
We use this expression to expand the radical of the expectation value of the volume operator (the square root of the operator in \eqref{simplified_volume}) --- this is a standard procedure in LQG \cite{Thiemann:2007zz}. The result is the square root of the eigenvalue in \eqref{simplified_volume}, which also coincides with the analogous expression in QRLG \cite{Alesci:2013xd}.

Finally, we can find the eigenequation of the U$(1)$ holonomy operators, derived on the basis states in \eqref{simplified_basis},
\begin{align}
\label{simplified_holonomy}
\hat{h}_{(p)}\big|l^{(p)}\!,\bar{m}^{(p)}\big>
=\exp\!\bigg(\tau^{(p)}\varepsilon\,\bar{c}_{(p)\!}-\i\,\bar{m}^{(p)}\!\!\int\!\!dx^q\bar{A}^{(p)}_q\bigg).
\end{align}
Notice that the reduction defined in sections \ref{V.2} and  \ref{V.3} is performed at the quantum level, hence we cannot \textit{a priori} replace the $\hat{h}_{(p)}^{(j)}$ operator with its eigenvalue $h_{(p)}^{(j)}$ in the Hamiltonian constraint. This operation can be done after we act with HCO on \eqref{simplified_basis}.

Then, in order to derive the action of HCO in our simplified approach, we find that the eigenvector of expression \eqref{loop} reads
\begin{align}
\label{simplified_loop}
\sum_{v\in\Gamma}\,\text{tr}\bigg(\hat{h}_{p\circlearrowleft q}\,\hat{h}_r^{-1}\Big[\hat{\mathbf{V}}_{\!v,l_0},\hat{h}_r\Big]\bigg)|\Gamma;J\rangle
=-\frac{\i\:\!\varepsilon}{2}\epsilon_{pqr}\sum_{v\in\Gamma}\,
\sin\!\big(\varepsilon\bar{c}_{(p)\!}\big)\sin\!\big(\varepsilon\bar{c}_{(q)\!}\big)
\frac{\sqrt{\kbar^3\prod_s^3\!\Big|\bar{m}^{s}(v)+\bar{m}^{s}\big(v-l_0^{s}\big)\Big|}}
{\bar{m}^{r}(v)+\bar{m}^{r}\big(v-l_0^{r}\big)}
I_v^{(r)}
|\Gamma;J\rangle\,,
\end{align}
with the inverse volume corrections in the form of
\begin{align}\label{simplified_corrections}
I_v^{(r)\!}=1+\frac{1}{8\Big(\bar{m}^{(r)}(v)+\bar{m}^{(r)}\big(v-l_0^{(r)}\big)\Big)^{\!\!2}}
+\mathcal{O}\Bigg(\frac{1}{ \big(\bar{m}^{(r)}\big)^{\!4}\! }\Bigg).
\end{align}
Here, by analogy with \eqref{simplified_matrix_volume}, we assumed that the eigenvalue of the square root of the volume operator square equals to the square root of the eigenvalue of $\big(\hat{\mathbf{V}}_{\!v,l_0}\big)^{\!2}$. Expanding the result in \eqref{simplified_loop} arbitrarily around $\bar{m}(v)\approx\bar{m}\big(v-l_0^{(s)}\big)\approx\infty$, leads to the well known eigenvalue of the Euclidean contribution to HCO in LQC \cite{Ashtekar:2003hd,Bojowald:2008zzb}. In order to derive the action of the operator in \eqref{curvature}, and then the full Lorentzian term, one should use the quantum relation introduced in \cite{Thiemann:1996aw}, namely
\begin{align}\label{simplified_curvature}
\hat{\text{K}}_v=-\frac{4}{\kbar^2}\epsilon^{pqr}
\bigg[\text{tr}\bigg(\hat{h}_{p\circlearrowleft q}\,\hat{h}_r^{-1}\Big[\hat{\mathbf{V}}_{\!v,l_0},\hat{h}_r\Big]\bigg),\hat{\mathbf{V}}_{\!v,l_0}\bigg]\,.
\end{align}
Details of the derivation of HCO for the cosmologically reduced LQG are in preparation \cite{Bilski_Lorentzian}. 


\section{Conclusions}\label{VII}

\noindent
We have argued that shortcomings arise within the initial model of Quantum Reduced Loop Gravity proposed by Alesci and Cianfrani.
We arrived to these conclusions re-examining the reduction procedure applied to the states of the kinematical Hilbert space of Loop Quantum Gravity, and developing a comparative analysis with previous attempts formulated in the literature of QRLG. Constraints were formerly inconsistently implemented within the framework of the reduced model, which was leading to an overconstrained dynamics, ill-defined Hilbert space and a Gribov obstruction.

This analysis motivated us to develop alternative implementations of the symmetry-reduction procedure, which we have discussed here in detail. Our findings are reminiscent of previous results in LQC. Nevertheless, our derivation, which aims at approaching the cosmological limit of LQG, reveals a novel attempt towards the minisuperspace quantization. We have been then shifting away from the paradigm of quantizing a symmetry reduced space, a procedure that so far (when considering the methodology of QRLG) could not help in solving the Hamiltonian constraint in the full theory. Instead, our proposal, following the standard gauge fixing methods, reckons on the program of bridging the gap between the full theory and former endeavors in LQC. However, the most important problem of any of the reduced variants of LQG, which is the question of whether the dynamics predictable by these modes is the same as the reduced dynamics of the full theory (not derivable due to its highly complicated formulation), remains open.

Finally, we leave as an open issue the question whether the reduction of the full quantum theory would lead to the same dynamics as the quantization of the classically reduced phase space. This has been however restricted to the single problem of the derivation of the Lorentzian sector of HCO.

The second open problem, not discussed in this article, refers to the idea to keep some symmetries of LQG unconstrained. This can be done, imposing gauge conditions that fix only the SU$(2)$ symmetry or the spatial diffeomorphisms. The latter issue can be investigated using the same gauge condition that we used here and that was originally proposed by Alesci, Cianfrani and Rovelli. As we demonstrated in Appendix \ref{E}, this condition Poisson-commutes with the Gauss constraint. In order to, conversely, freeze the Gauss constraint, leaving the spatial diffeomorphisms unmodified, one needs to construct another condition, which will commute with the vector constraint.
\\


\appendix

\section{$\mathfrak{su}(2)$ representations and spin representations}\label{A}

\noindent
The Pauli matrices $\sigma^i$ are a set of three $2\times2$ complex matrices. They are Hermitian (self-adjoint), thus they represent observables in quantum theories. They are also unitary, hence they preserve norm and thereby probability amplitude. They satisfy normal commutation relations,
\begin{align}\label{sigma}
[\sigma^i,\sigma^j]=2\i\epsilon^{ijk}\sigma^k\,.
\end{align}

The $\mathfrak{su}(2)$ Lie algebra generators $t^i:=-i\sigma^i$ of the SU$(2)$ group are antihermitian (skew-Hermitian) and unitary. They form $\mathfrak{su}(2)$ basis and satisfy the following commutation relations,
\begin{align}\label{Tmatrix}
[t^i,t^j]=2\epsilon^{ijk}t^k\,.
\end{align}

The $\mathfrak{so}(3)\cong\mathfrak{su}(2)$ Lie algebra generators $\big\{u^1\!:=\!\frac{\i}{\sqrt{2}}(\sigma^1\!+\!i\sigma^2),\,u^2\!:=\!-\frac{\i}{\sqrt{2}}(\sigma^1\!-\!i\sigma^2),\,u^3\!:=\!-\i\sigma^3\big\}$ of the rotation group are antihermitian and unitary. They form spherical basis and satisfy the following commutation relations,
\begin{align}\label{Umatrix}
[u^1,u^2]=2\i u^3\,,\ [u^1,u^3]=2\i u^1\,,\ [u^2,u^3]=-2\i u^2\,.
\end{align}

The Hermitian generators of the spin representation in particle physics are defined as $s^i:=\frac{1}{2}\sigma^i$.

The antihermitian generators of $\mathfrak{su}(2)$ that form the spin representation in LQG are defined as $\tau^i:=\frac{1}{2}t^i$. They are the preferable choice for keeping both holonomies and Ashtekar connections real.

Finally, we should define the antihermitian generators of the $\mathfrak{so}(3)$ spherical representation of spin in LQG, $\upsilon^i=\frac{1}{2}u^i$. Knowing the standard spin basis (equivalently the angular momentum basis) defined by the $\i\upsilon^i$ operators (commonly used in the particle physics), we find
\begin{align}
\begin{split}
\label{upsilon123}
\upsilon^1\big|j,m\big>=&\,-\i\sqrt{j(j+1)-m(m+1)}\big|j,m+1\big>\,,
\\
\upsilon^2\big|j,m\big>=&\,-\i\sqrt{j(j+1)-m(m-1)}\big|j,m-1\big>\,,
\\
\upsilon^3\big|j,m\big>=&\,-\i m\big|j,m\big>\,,
\\&
\end{split}
\\
\label{upsilonSquare}
\upsilon^i\upsilon^i\big|j,m\big>=&\,-j(j+1)\big|j,m\big>\,,
\end{align}
where $j=0,1/2,1,...$ and $m=-j,-j+1,...,j$.

It is worth noting that all the representations discussed above are proper. Furthermore, the sign in front of each triple of generators is conventional. Reversing the sign, we impose anomalous commutation relations instead of the normal ones in \eqref{sigma} and \eqref{Tmatrix}.


\section{Wigner $D$-matrices of diagonal holonomies}\label{B}

\noindent
Let us define a unitary rotation matrix $\vec{\rho}$ as follows,
\begin{align}
\rho^1=&\frac{1}{\sqrt{2}}\left(\begin{array}{cc}1&\!-1\\1&1\end{array}\right),
\quad&
\rho^2=&\frac{1}{\sqrt{2}}\left(\begin{array}{cc}1&\i\\\i&1\end{array}\right),
\quad&
\rho^3=&\left(\begin{array}{cc}1&0\\0&1\end{array}\right).
\end{align}
Applying SU$(2)$ covariance of generators $\tau^i$, we can consider the rotation of a basis frame as a matrix operation,
\begin{align}
e^{\vec{\rho}\cdot\vec{\tau}\,\varepsilon\,\bar{c}}=\rho^{\dagger}e^{\tau^3\varepsilon\,\bar{c}}\rho\,.
\end{align}
Then the following relation holds,
\begin{align}
D^{(j)}_{mn}\big(e^{\vec{\rho}\cdot\vec{\tau}\,\varepsilon\,\bar{c}}\big)
=D^{(j)}_{mm'}\big(\rho^{\dagger}\big)D^{(j)}_{m'n'}\big(e^{\tau^3c}\big)D^{(j)}_{n'n}\big(\rho\big)
=e^{-\i mc}\,,
\end{align}
where we used the fact that for a diagonal SU$(2)$ element, the Wigner $D$-matrix \cite{Wigner} takes a particularly simple form,
\begin{align}
D^{(j)}_{mn}\big(e^{\tau^3c}\big)=e^{-\i mc}\delta_{mn}\,.
\end{align}


\section{Symplectic structure and Poisson brackets}\label{C}

\noindent
The symplectic structure of LQG, $\Omega_{LQG}(\delta_1,\delta_2)$, corresponding to \eqref{Poisson_LQG} reads
\begin{align}
\Omega_{LQG}=&\;\frac{2}{\gamma\kappa}
\!\int\!\!d^3x\Big(\delta_1A^i_a(x)\,\delta_2E_i^a(x)-\delta_2A^i_a(x)\,\delta_1E_i^a(x)\Big)\,.
\intertext{Thus the analogous structure for QRLG becomes}
\label{symplectic_QRLG}
\Omega_{QRLG}=&\;\frac{2}{\gamma\kappa}\sum_i^3\!
\int\!\frac{d^3x}{{\bf V}_{\!0}}\Big(
\delta_1c_{(i)}\big(t,x^{(i)}\big)\,\delta_2p^{(i)}\big(t,x^{(i)}\big)-\delta_2c_{(i)}\big(t,x^{(i)}\big)\,\delta_1p^{(i)}\big(t,x^{(i)}\big)
\Big)\,,
\intertext{while considering the LQC limit, one finds}
\Omega_{LQC}=&\;\frac{2}{\gamma\kappa}\sum_i^3d\bar{c}_{(i)}(t)\wedge d\bar{p}^{(i)}(t)
=\frac{2}{\gamma\kappa}\,d\bar{c}_{i}(t)\wedge d\bar{p}^{i}(t)\,.
\end{align}
Therefore the Poisson bracket \eqref{Poisson_LQG} for the reduced variables in \eqref{R-C_A} and \eqref{R-C_E} reads
\begin{align}
\label{Poisson_QRLG}
\big\{c_{i}\big(t,x^{(i)}\big),p^{j}\big(t,y^{(j)}\big)\big\}=&\;\frac{\kappa\gamma}{2}\delta_i^j\delta\big(x^{(i)}-y^{(j)}\big)
=\frac{\kbar}{\hbar}\delta_i^j\delta\big(x^{(i)}-y^{(i)}\big)\,.
\end{align}
Analogously, in the homogeneous limit one finds
\begin{align}
\label{Poisson_LQC}
\{\bar{c}_{i}(t),\bar{p}^{j}(t)\}=&\;\frac{\kappa\gamma}{2}\delta_i^j=\frac{\kbar}{\hbar}\delta_i^j\,.
\end{align}


\section{Standard gauge fixing methods}\label{D}

\noindent
In this Appendix we recall the essence of the standard operator methods of quantization described in details in \cite{Henneaux:1992ig}. For the sake of simplicity, which is nonetheless sufficient for our purposes, we assume a system of only bosonic fields. All the statements written below without a proof, are verified in the cited textbook.

\subsection{Classical fixation}\label{D.1}

\noindent
Before analyzing quantum methods applied to simplified gauge systems, let us sketch briefly how a simplified procedure can be achieved already at the classical level. Formally, one should define \textit{ad hoc} restrictions, called canonical gauge conditions, in the form of a set of equations. For simplicity, we discuss only the set of independent equations, while the generalization is straightforward.

Let us recall two facts. The number of gauge conditions must be equal to the number of (also independent) first class constraints that are fixed. The gauge conditions and corresponding first class constraints form a second class set. Therefore after fixing the gauge, the theory becomes restricted to a constrained surface of the phase space, and one should then use Dirac brackets instead of Poisson brackets.

There is, however, a possible problem that could arise while the gauge conditions are imposed. These are solved locally and may not necessarily intersect the gauge orbits only once. This issue is called a Gribov obstruction and its absence should be considered as another condition for a globally well-defined form of a set of the gauge conditions.

Another problem is the quantization of the second class constraints, we will discuss that in the next subsections. In many cases, however the set of second class constraints can be graded down to a first class system, either introducing gauge conditions or enlarging the set of canonical variables, thus introducing other degrees of freedom and the corresponding symmetries. This method is not unique, but it controls the number of the overall propagating degrees of freedom. Thus, starting with a system of few first class constraints, and then introducing gauge conditions, one arrives to a set of second class constraints. Nonetheless, one may return either to another system having a higher number of first class constraints or to a different set of first class constraints, endowed with different canonical gauge conditions. This classical variety could lead, however, to different quantum systems, thus later to different physics --- for instance different corrections in a perturbative semiclassical analysis.
\\

A method to get around these troubles, rather than solving them, is to modify the classical action (or the total Hamiltonian). In physics, one is often interested only in some aspects of the whole theory, which can be identified as a new model obtained by restricting the general theory to a simpler one, while preserving higher symmetry by construction (not as a gauge symmetry). After this change of the model, even if this still describes the same number of degrees of freedom as the gauge-fixed general theory, one cannot be sure that the dynamics of the `smaller' model coincides with the `bigger' one on the constrained surface. These theories are not connected, hence to prove that the `smaller' theory with a higher symmetry corresponds to the `bigger' with solved constraints, requires anyhow to quantize the system of constraints.

\subsection{Correspondence rules of quantization}\label{D.2}

\noindent
The easiest method of quantization of the second class constraints is to redefine the rules of canonical quantization, setting these constraints to zero operators. This way, one obtains a quantum theory from a lower dimensional phase space. If the result corresponds to the classical symmetry that is enough to describe a phenomenon one wants to investigate, one obtains a quantum reduced system similar to the classical fixation, described in \hyperref[D.1]{D.1}. The advantage with respect to the quantization of the classically reduced system stands in the presence of the original constraint operator equations, restricted now to the lower-dimensional operators. As long as the constraints are not equal (up to a constant) to the canonical variables, the quantum reduction, which preserves the original structure of the operator equations, is a significantly more general method. The actions of quantum-reduced operators become restricted to coincide with the states without the removed symmetry. The classical fixation conversely does not restrict canonical quantization rules to set some variables to zero operators, but it directly removes these variables. This way, the classical fixation changes the theory into another one, while the more general method of the correspondence rules of quantization selects the sectors of operators, which would be non-vanishing while solving the second class constraints.

To complete the description of this procedure, we should remind that quantization is now defined by replacing Dirac brackets by commutators times $1/\i\hbar$. The problem may occur in finding a quantum representation for different solutions of the second class constraints --- only in this case Dirac brackets are not (weakly) equal to the Poisson brackets. If these solutions either do not form any algebra or are not simply numbers, one may not be able to solve this issue.

A better option could be then to introduce extra degrees of freedom in terms of new canonical variables. This then turns to be a problem of first class constraints quantization, just for a different, but larger, number of constraints, which can be quantized as any other first class system.

\subsection{Reduced phase space quantization}\label{D.3}

\noindent
The reduced phase space quantization is a method of quantization of gauge-invariant functions restricted to the equivalence classes of gauge orbits. This method appears to be the most natural one, since after the reduction, one uses only the regular Poisson brackets. The problem here is to find \textit{a priori} a complete set of gauge-invariant functions, which will be quantized. This can be a achieved easily, as long as the Gribov obstruction is not present. In that case, one simply imposes globally gauge fixing conditions. In other words, any reduced functional of the canonical variables must coincide with an appropriate restriction on the corresponding gauge invariant functional. Gauge invariant functions are transformed then into operators, while Dirac brackets, restricted by the gauge conditions to the reduced phase space, turn into Poisson brackets. The canonical quantization is then straightforward.

The problem of this method is that removing gauge invariance may spoil a crucial symmetry. For instance, a classically reduced theory could be still perturbatively expanded around the reduced symmetry, in terms of the reduced variables. The quantum reduced theory does not contain states allowing on this expansion, hence it has to remain permanently symmetry-fixed both at the purely quantum level, as well as while deriving the semiclassical solutions. Another difficulty is that this method destroys locality of the phase space.


\section{Algebra of gauge conditions}\label{E}

\noindent
We analyzed in this paper the reduced phase space quantization obtained by fixing the canonical gauge conditions, $\chi_i\approx0$ and $\eta^{ab}\approx0$, which are defined as,
\begin{align}
\label{chi}
\chi_i:=&\;\epsilon_{ijk}{}^{0\!}E^j_k\,,
\intertext{and}
\label{eta}
\eta^{ab}:=&\;E^a_iE^b_i\big|_{a\neq b}\,,
\end{align}
respectively. The densitized dreibein projected along fiducial directions, introduced above, reads
\begin{align}
\label{projected_E}
{}^{0\!}E^i_j:={}^{0\!}e^i_aE^a_j\,.
\end{align}
Notice that the formula for the reverse fiducial densitized dreibien is not denoted as usually in the literature \cite{Thiemann:2007zz}, thus $\big(E^i_j\big)^{\!-1}\!\neq E^j_i$.

In this appendix we present the algebra between gauge conditions and all the constraints, being elements of the Hamiltonian in \eqref{A_total_Lambda}. In what follows, we need to derive the following functional derivatives,
\begin{align}
\frac{\delta\chi_i}{\delta E^c_j}=-\epsilon_{ijk}{}^{0\!}e^k_c\,,
\qquad
\frac{\delta\eta^{ab}}{\delta E^c_j}=\big(E^a_j\delta^b_c+E^b_j\delta^a_c\big)\!\big|_{a\neq b}\,,
\qquad
\frac{\delta\chi_i}{\delta A_c^j}=\frac{\delta\eta^{ab}}{\delta A_c^j}=0\,.
\end{align}
It is worth mentioning that, for the purpose of the analysis in Sec.~\ref{IV}, we should consider the square of the $\chi_i$ condition. This however, is not going to significantly change any result below, since the equation $\delta\chi^2/\delta E^c_j=-2\chi_i\epsilon_{ijk}{}^{0\!}e^k_c$ is proportional to \eqref{chi}.

Let us first verify that the gauge conditions are independent, indeed finding
\begin{align}
\label{chi_eta}
\big\{\chi_i,\eta^{ab}\big\}=0\,.
\end{align}
As a next step, we should derive Poisson brackets between both gauge conditions and all the constrains. For simplicity, we neglect the integrals, and the Lagrange multipliers in the constraints. These latter would vanish anyway integrating out the Dirac deltas, and not influence the results, carrying no additional degrees of freedom. We restrict then the Poisson bracket equations below to constraint densities.

Let us first solve the auxiliary equations, containing all the non-vanishing functional derivatives, finding
\begin{align}
\frac{\delta\mathcal{G}_i^{(A)}\!\!\!}{\delta A_c^j}=&\;\frac{1}{\gamma\kappa}\epsilon_{ijk}E^c_k\,,
\\
\frac{\delta\mathcal{V}_a^{(A)}\!\!\!}{\delta A_c^j}=&\;\frac{1}{\gamma\kappa}\big(\delta^c_a\delta^b_d-\delta^b_a\delta^c_d\big)D_bE^d_j\,,
\\
\frac{\delta\big(\mathcal{H}^{(A)}\!+\mathcal{H}^{(\Lambda)}\!\big)\!\!}{\delta A_c^j}
=&\;\frac{2}{\kappa\sqrt{q}}\big(\delta^c_a\delta^b_d-\delta^b_a\delta^c_d\big)E^a_k\bigg(
\frac{1}{\gamma^2}A^k_bE^d_j+\epsilon_{jkl}\partial_bE^d_l
\bigg)\,.
\end{align}
It is easy then to derive the Poisson algebra between all the constraints and the gauge conditions $\chi_i$ and $\eta^{ab}$, getting
\begin{align}
\label{Gauss_chi}
\Big\{\mathcal{G}_i^{(A)}\!,\chi_j\Big\}=&\;\frac{1}{\gamma\kappa}\big({}^{0\!}E^i_j-{}^{0\!}E^k_k\delta^i_j\big)\,,
\\
\label{vector_chi}
\Big\{\mathcal{V}_a^{(A)}\!,\chi_i\Big\}=&\;\frac{1}{\gamma\kappa}\epsilon_{ijk}\big({}^{0\!}e^k_aD_bE^b_j-D_a{}^{0\!}E^k_j\big)\,,
\\
\label{scalar_chi}
\Big\{\mathcal{H}^{(A)}\!+\mathcal{H}^{(\Lambda)}\!,\chi_i\Big\}
=&\;\frac{2}{\kappa\sqrt{q}}\bigg(
\frac{1}{\gamma^2}\epsilon_{ijk}\big({}^{0\!}E^j_lE^a_k-{}^{0\!}E^j_kE^a_l\big)A^l_a
+\partial_a\big({}^{0\!}E^j_iE^a_j-{}^{0\!}E^j_jE^a_i\big)
\bigg)\,,
\intertext{and}
\label{Gauss_eta}
\Big\{\mathcal{G}_i^{(A)}\!,\eta^{ab}\Big\}=&\;0\,,
\\
\label{vector_eta}
\Big\{\mathcal{V}_a^{(A)}\!,\eta^{bc}\Big\}=&\;\frac{1}{\gamma\kappa}\Big[
E^b_j\big(D_dE^d_j\delta_a^c-D_aE^c_j\big)
+E^c_j\big(D_dE^d_j\delta_a^b-D_aE^b_j\big)
\Big]\!\Big|_{b\neq c}\,,
\\
\label{scalar_eta}
\Big\{\mathcal{H}^{(A)}\!+\mathcal{H}^{(\Lambda)}\!,\eta^{ab}\Big\}
=&\;\frac{2}{\kappa}\bigg(
\frac{\sqrt{q}}{\gamma^2}A^i_c\big(q^{ac}E^b_i+q^{bc}E^a_i-2q^{ab}E^c_i\big)
+\frac{1}{\sqrt{q}}\epsilon_{ijk}E^c_i\big(E^a_j\partial_cE^b_k-E^b_k\partial_cE^a_j\big)
\bigg)\!\bigg|_{a\neq b}\,.
\end{align}


\vspace{1cm}

{\it{\textbf{Acknowledgments}}}\\
\noindent
A.M. acknowledges support by the NSFC, through the grant No. 11875113, the Shanghai Municipality, through the grant No. KBH1512299, and by Fudan University, through the grant No. JJH1512105. J.B. is supported in part by the NSFC, through the grants No. 11375153 and 11675145.



\end{document}